\documentclass[%
 reprint,
nofootinbib,
 amsmath,amssymb,
 aps,
]{revtex4-2}

\usepackage{graphicx}
\usepackage{dcolumn}
\usepackage{bm}

\begin{document}


\title{Equilibration and ``Thermalization'' in the Adapted Caldeira-Leggett model}

\author{Andreas Albrecht}
\affiliation{%
Center for Quantum Mathematics and Physics and Department of Physics and Astronomy\\
University of California, Davis\\
One Shields Avenue\\ Davis, CA 95616
}%





\begin{abstract}
I explore the processes of equilibration exhibited by the Adapted Caldeira-Leggett (ACL) model, a small unitary ``toy model'' developed for numerical studies of quantum decoherence between an SHO and an environment.  I demonstrate how dephasing allows equilibration to occur in a wide variety of situations.  While the finite model size and other ``unphysical'' aspects prevent the notions of temperature and thermalization from being generally applicable, certain primitive aspects of thermalization can be realized for particular parameter values.  I link the observed behaviors to intrinsic properties of the global energy eigenstates, and argue that the phenomena I observe contain elements which might be key ingredients that lead to ergodic behavior in larger more realistic systems.  The motivations for this work range from curiosity about phenomena observed in earlier calculations with the ACL model, to much larger questions related to the nature of equilibrium, thermalization and the emergence of physical laws. 
\end{abstract}

\maketitle

\section{Introduction}
\label{sec:intro}
In~\cite{ACLintro}, my collaborators and I introduced a toy model which adapted the Caldeira--Leggett model for numerical analysis. This ``Adapted Caldeira--Leggett'' (ACL) model was designed to optimize decoherence and einselection between a simple harmonic oscillator (SHO) and an environment. The SHO and environment are treated together as a closed, unitarily evolving quantum system. The ACL model naturally equilibrates when evolved for a sufficient length of time.  We used equilibrium states thus obtained in~\cite{ACLeqm} to study the extent to which equilibrium states can exhibit einselection~\cite{Zurek:1992mv} despite the absence of an arrow of time. That work was motivated especially by cosmological considerations. 

In this paper, I dig deeper into the equilibrium behavior exhibited by the ACL model. I show that a range of equilibration behaviors is possible, some of which show attributes that might be seen as a primitive form of ``thermalization,'' and others that definitely do not.  I show how the different behaviors are controlled by parameters in the ACL Hamiltonian and relate these behaviors to properties of the global energy eigenstates. 

One motivation for this work is to make sure the equilibrium states of the ACL model form a suitable foundation for the studies in~\cite{ACLeqm} (they do). However, by design, the ACL model eschews a number of physically realistic features (such as locality in the environment) to allow decoherence to function efficiently with limited computational resources. One might ask, what is the point of studying equilibration in a system with significant unphysical features?  While in~\cite{ACLintro}, and also in~\cite{CopyCat}, we do make some connection to results from Nuclear Magnetic Resonance experiments, for the purposes of this paper, I regard the physically unrealistic aspects of the ACL model as a strength. 

I have long been fascinated by the question of the emergence of physical laws through the identification of (possibly multiple) semiclassical domains in large quantum systems (as discussed, for example, in~\cite{Albrecht:2007mm,AlbrechtPhysRevD.91.043529}). One feature that seems to be important is the capability of large numbers of degrees of freedom to behave in very simple ways that have a semiclassical description.  In physically realistic situations, this is often achieved by the processes of equilibration and thermalization.  I am interested in turning the question around and learning in general terms what sorts of physical systems can achieve these behaviors.  I am curious if the need to have equilibration and thermalization can help choose---in some selection process associated with their emergence---key features of the laws of physics as we know them. This paper takes a very small step in that direction by exploring the various behaviors of the ACL model.

I introduce the ACL model in Section~\ref{sec:ACL} and in Section~\ref{sec:basic} demonstrate how the process of dephasing lies at the root of the wide range of equilibration processes that the model exhibits. I point out that, quite generically, the dephasing processes provide all that is needed to produce suitable equilibrium states for the work in~\cite{ACLeqm}. In Section~\ref{sec:EnoT}, I show that the processes depicted in Section~\ref{sec:basic} are missing a key feature associated with thermalization. Sections~\ref{sec:varyEI} scan a range of results produced by varying a parameter in the Hamiltonian. I show how an appropriate choice of this parameter allows the ACL model to include the thermalizing features, and argue that additional approximately conserved quantities are present in cases in which such features are absent.  The discussion up to Section~\ref{sec:varyEI} has been shaped by tracking the energies of the SHO and the environment, specifically the first moments of the energy distributions for each of these subsystems. Section~\ref{sec:edist} expands the discussion to scrutinize the full energy distributions for these subsystems, starting with their initial forms and tracking them as they settle into equilibrium. I explore how these also depend on the parameters of the system and study the presence or absence of the thermalizing behavior in this context.  Section~\ref{sec:w_eigenstates} examines the properties of the global energy eigenstates, first establishing their general properties and then relating those to various behaviors reported earlier in the paper. The question of the tuning of parameters and the initial states is examined briefly in Section~\ref{sec:tuning}, and Section~\ref{sec:dc} presents some further discussion and conclusions. Two appendices address various technical details, and a third explores the relationship of my ACL calculations to the eigenstate thermalization hypothesis.  This is an invited paper for a special volume honoring the 70th birthday of Wojciech Zurek, and I offer some appropriate reflections in Section~\ref{sec:refl}.

\section{ACL Model}
\label{sec:ACL}
In~\cite{ACLintro}, my collaborators and I introduced a toy model which adapted the Caldeira--Leggett model for numerical analysis. This ``Adapted Caldeira--Leggett'' (ACL) model has a ``world'' Hamiltonian  $H_w$ of the form
\begin{equation}
H_w = H_s \otimes {\bf{1}}_{}^e + H^I + {{\bf{1}}^s} \otimes {H_e}
\label{eqn:Hform}
\end{equation}
where the ``system'' Hamiltonian $H_s$ represents a truncated simple harmonic oscillator~(SHO).

The interaction term is given by $H^I = {q_{s}}\otimes H_e^I$, where $q_s$ is the SHO position operator and $H_e^I$ has the form 
\begin{equation}
H_e^I = {E_I}R_I^e + E^0_I.
\label{eqn:Hiedef}
\end{equation}
The matrix $R_I^e$ is a random matrix constructed by drawing each of the real and imaginary parts of each independent matrix element of this $N_e\times N_e$ Hermitian matrix from a distribution that is uniform over the interval $[-0.5,0.5]$.   Throughout this paper, I will use $N_e=600$ (the size of the environment Hilbert space) and $N_s=30$ for the truncated SHO (as used in~\cite{ACLintro,ACLeqm}).

The environment self-Hamiltonian is given by 
\begin{equation}
H_e = {E_e}R^e + E^0_e
\label{eqn:Hedef}
\end{equation}
where $R^e$ is constructed in the same manner as $R_I^e$, but as a separate realization. In \mbox{Equations~\eqref{eqn:Hiedef} and \eqref{eqn:Hedef}}, $E_I$ and $E_e$ are c-numbers which parameterize the overall energy scales. Both $R_I^e$ and $R^e$ are fixed initially and are not changed during the time evolution. The full Hamiltonian of the ACL model is time independent. All the results in this paper use $E^0_I=E^0_e=0$, but nonzero values for these offset parameters have been useful in other contexts such as~\cite{ACLeqm}. 

In~\cite{ACLintro}, we demonstrated how the ACL model is able to numerically reproduce decoherence phenomena typically studied with the original Caldeira--Leggett model, and argued that the specific form of $H_w$ enables the numerical studies to reproduce these phenomena in an efficient manner.  In~\cite{ACLeqm}, we used ACL model calculations to address the relationship between the arrow of time and the emergence of classicality (a topic we motivated with cosmological considerations), and, in~\cite{CopyCat}, we explored new phenomena at the early stages of decoherence. 

We made a point in~\cite{ACLintro} of demonstrating the capability of the ACL model to equilibrate, and these equilibrated states played a key role in~\cite{ACLeqm}.  The focus of this paper is to more fully understand the equilibration processes in the ACL model, and place them in the context of modern ideas from quantum statistical mechanics. 

\section{Basic Equilibration and Dephasing}
\label{sec:basic}

\begin{figure}[h]
\includegraphics[width=3.7 in]{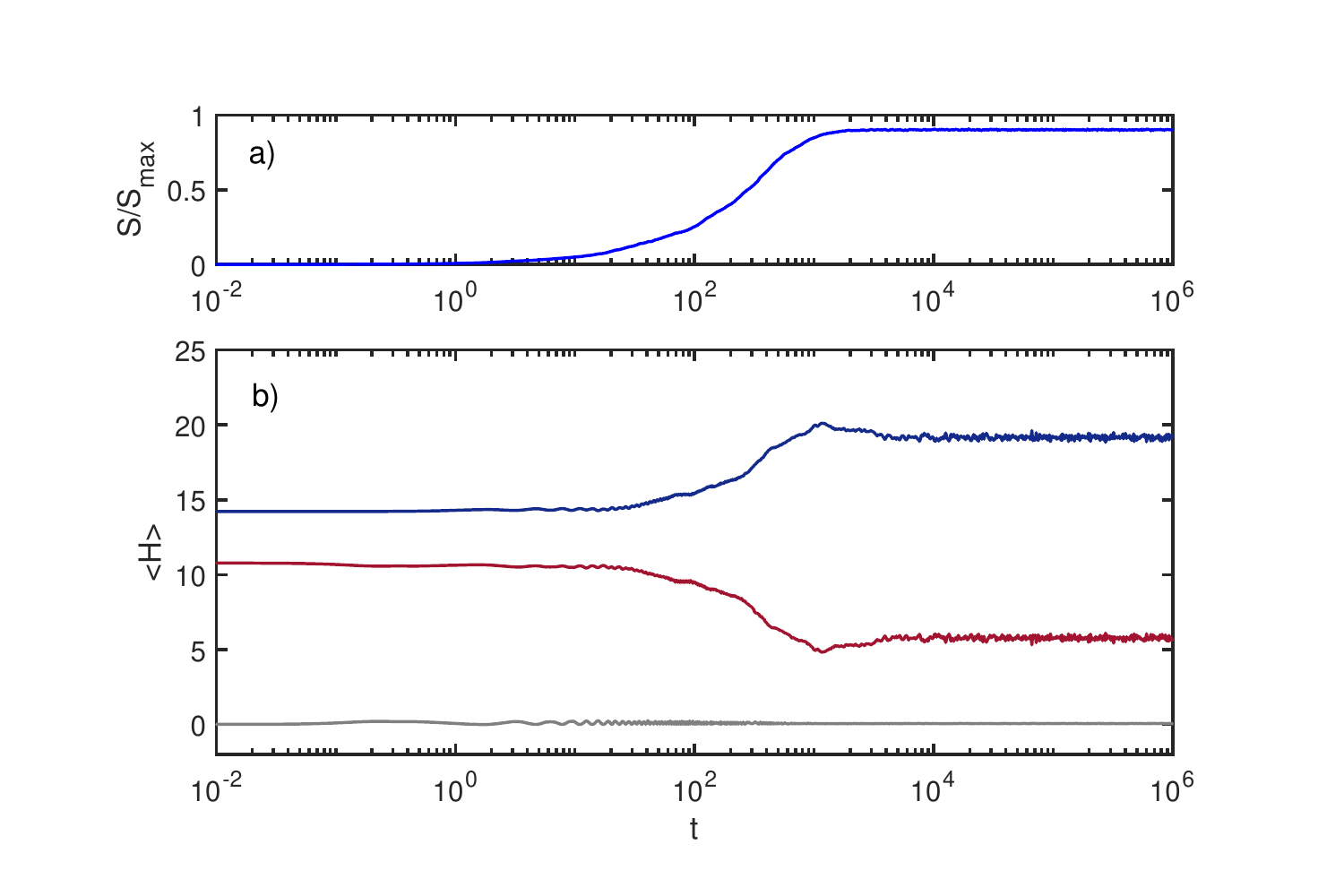}
\caption{{Equilibration in the ACL model:} 
  Entropy increases and energy flows from the environment to the SHO for a period of time, and then stabilizes up to small fluctuations. (\textbf{a}) entanglement entropy between the SHO and the environment; (\textbf{b}) subsystem energies $\left<H_s\right>$ (blue) and $\left<H_e\right>$ (red), and interaction energy $\left<H^I\right>$ (grey).
\label{fig:FirstFlow}}
\end{figure}

The basic equilibration process of the ACL model is demonstrated in Figure~\ref{fig:FirstFlow}. In this example, energy flows from the SHO to the environment for a period of time, and then the energies in both systems stabilize, up to small fluctuations. 
  For these curves, I used $E_e=1$ and $E_I=0.02$. Throughout this paper, I use units where $\hbar \omega_{SHO} = 1$. The initial state is a product of a coherent state for the SHO and an eigenstate of $H_e$, each with energies as shown in the plot. 

\begin{figure}[h]
\includegraphics[width=3.7 in]{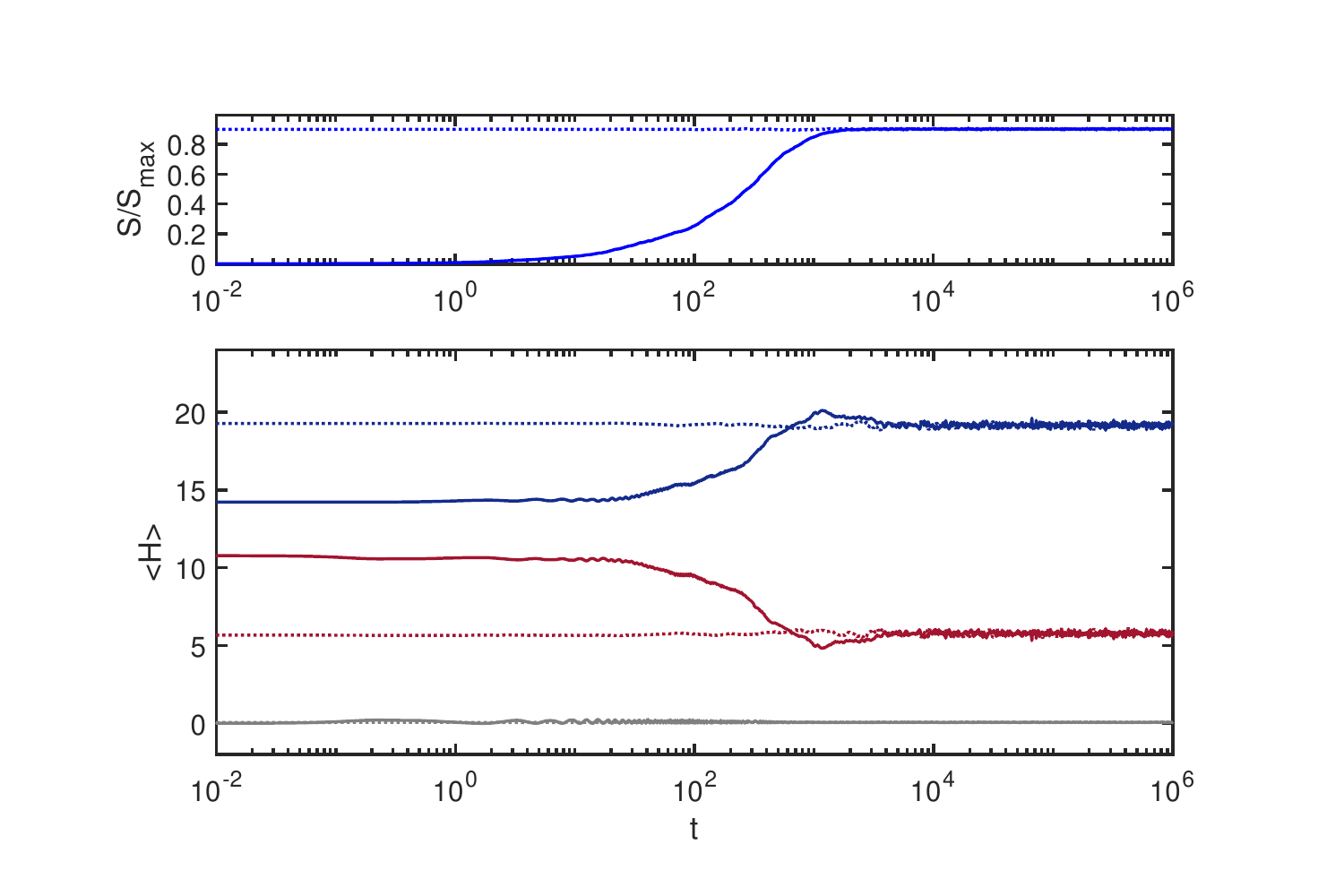}
\caption{{Dephasing: In addition}  
 to the curves shown in Figure~\ref{fig:FirstFlow}, I have added dotted curves from the calculations with randomized $\alpha_w^i$ phases discussed in the text. The convergence of the dotted and solid curves at later times reflects the dephasing nature of the equilibration process. 
\label{fig:FirstFlowWithRandom}}
\end{figure}

As inferred in Appendix A of~\cite{ACLeqm}, the basic mechanism for this equilibration is ``dephasing.''  When expanding the global state as
\begin{equation}
    \left|\psi_w\right> = \sum_i \alpha_w^i(t)\left|E_w^i\right>
\end{equation}
where $\left|E_w^i\right>$ are the eigenstates of $H_w$, special relationships are required among the $\alpha_w^i(t=0)$ to realize the initial product form of $\left|\psi_w\right>$.  The dimension of the global space ($N_w=N_s\times N_e=$ 18,000) is sufficiently large, and the eigenvalues of $H_w$ are sufficiently incommensurate that these special relationships come undone over time. The equilibrium state corresponds to the state where 
the phases of the $\alpha_w^i(t=0)$ are fully randomized. This is demonstrated explicitly in Figure~\ref{fig:FirstFlowWithRandom} where additional curves are included from an initial state where the phases of $\alpha_w^i$ were randomized ``by hand'' at $t=0$.  

This dephasing process is well-known in the quantum statistical mechanics literature (as reviewed, for example, in~\cite{binder2019thermodynamics}. {Some nice} historical reflections can be found in~\cite{Lloyd2006ExcuseOI}).  
Papers such as~\cite{CPZ1,CPZ2} also demonstrate its general relevance to decoherence. In Appendix~\ref{app:DDD} I elaborate a bit on how I used the term dephasing here, and its connection to other topics such as decoherence. 

\begin{figure}[h]
\includegraphics[width=3.7 in]{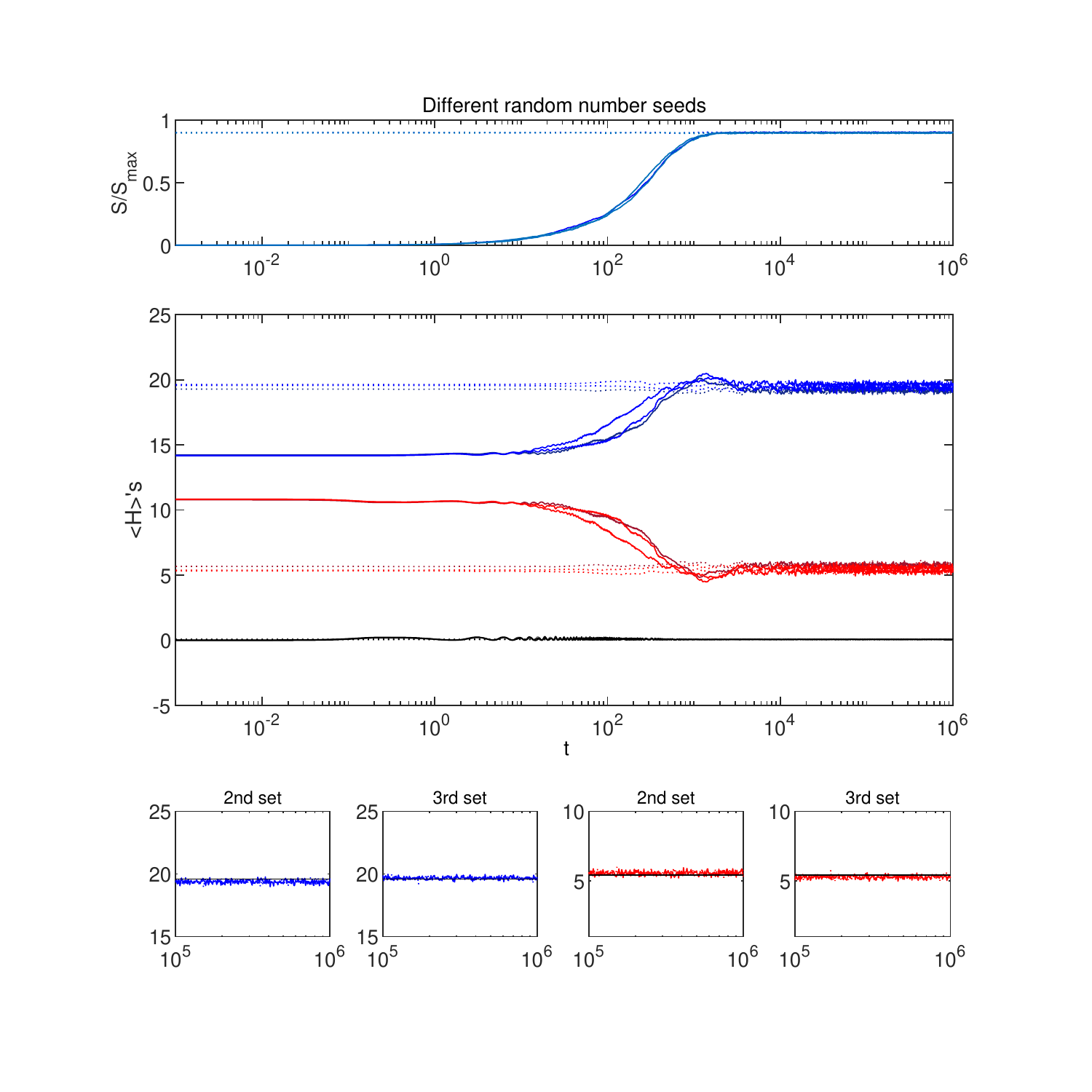}
\caption{{Random artifacts: } 
The curves from Figure~\ref{fig:FirstFlowWithRandom} are shown with additional curves (in brighter shades) giving the same calculations using different random number seeds.  The small panels zoom in on the $\left< H \right>$'s during equilibrium, and show the same state in each panel, with and without the randomized phases. The matched pairs converge tightly (supporting the dephasing picture), while subtle changes to the Hamiltonian from changing the random number seed generate a slightly larger scatter from one set to another. The small panels have black horizontal lines that are the same for both sets as a reference to aid in tracking the vertical scatter between the sets.   
\label{fig:FirstFlowMoreRands}}
\end{figure}

Figure~\ref{fig:FirstFlowMoreRands} shows the effect of choosing different random number seeds on curves from Figure~\ref{fig:FirstFlowWithRandom}. In this figure, the curves from Figure~\ref{fig:FirstFlowWithRandom} are reproduced along with four additional curves. The additional curves were generated the same way, except with different seeds for the random number generator used for randomizing phases and for the random entries in $H_e$ and $H_e^I$.  The similarity of these sets of curves reflects the fact that artifacts of individual random number seeds show up only in the small scale fluctuations. 

I note that the equilibration process driven by dephasing presented here is more than sufficient as a basis for the equilibrium states studied in~\cite{ACLeqm}. For that work, the crucial piece is that detailed balance should be respected, so that fluctuations and their time reverse are equally likely to appear.  Next, I turn to more nuanced aspects of equilibration in this model which I find interesting, although not specifically in the context of~\cite{ACLeqm}. 
 
\section{Equilibration without ``Thermalization''}
\label{sec:EnoT}
\begin{figure}[h]
\includegraphics[width=3.7 in]{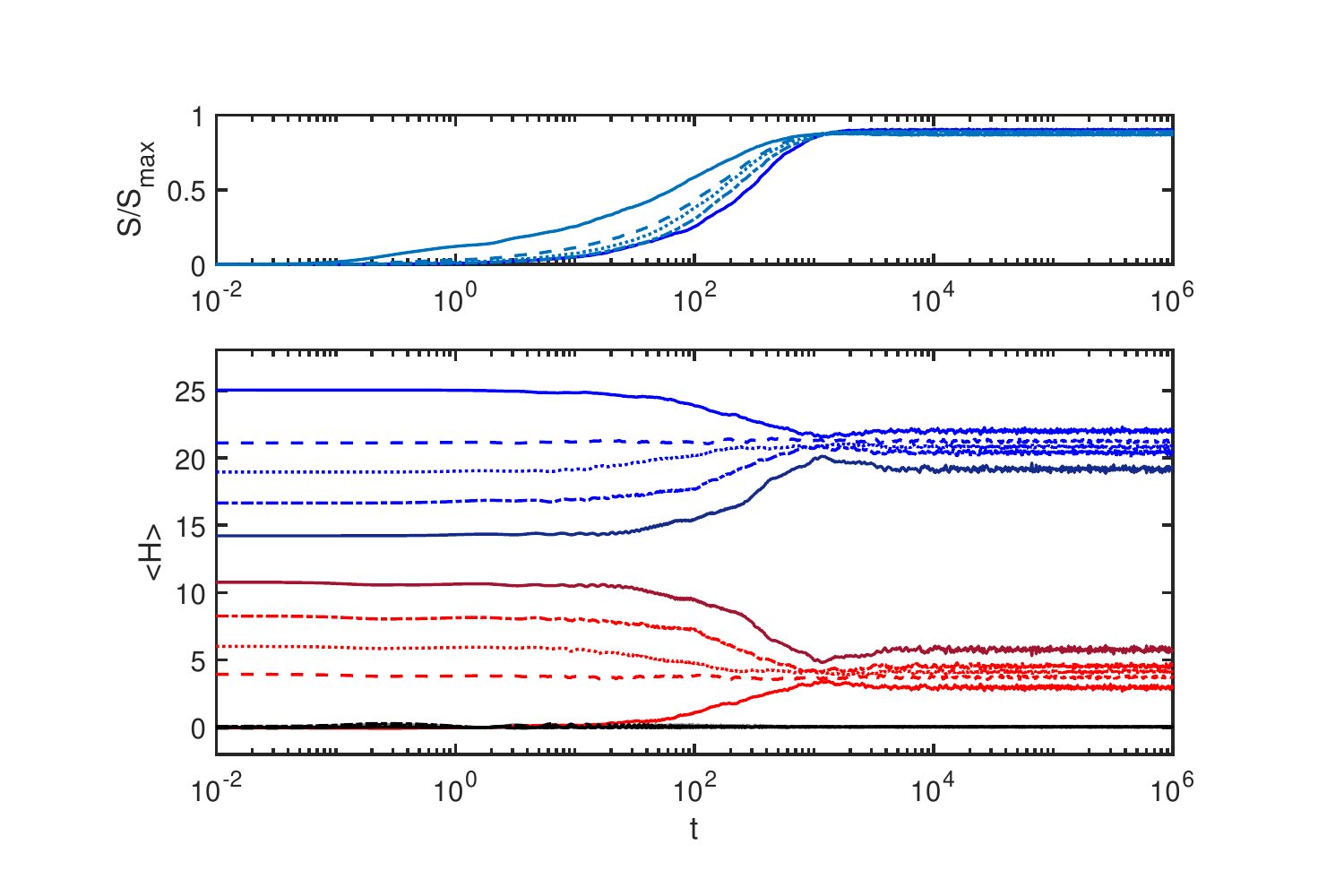}
\caption{{Varying the initial conditions:} 
 In addition to the curves from Figure~\ref{fig:FirstFlow}, I have added results from different initial conditions (all with the same total energy $\left<H_w\right>$). The new curves have a different shade of the same color and each set ($S$, $\left<H_s\right>$ and $\left<H_e\right>$) is matched by line type. Note that, while the total energy is the same, at the end of the equilibration process, the energy distribution between the environment and the SHO is different in each case. Here, $E_I=0.02$.
\label{fig:AllFlows020}}
\end{figure}

Figure~\ref{fig:AllFlows020} plots the entanglement entropy and subsystem energies from Figure~\ref{fig:FirstFlow} along with equivalent curves produced with different initial conditions.  
 In each case, the total energy was set at $\left<H_w\right>=25$, but the initial energy was distributed differently between the environment and the SHO.  As with Figure~\ref{fig:FirstFlow}, all SHO initial states were coherent states with initial energies as shown in the plot, and the environment initial states were eigenstates of $H_e$.  More details about how the initial conditions are constructed appear in Appendix~\ref{app:IC}.  

One expects isolated physical systems with the same global energy to thermalize to the same distribution of energies among subsystems, regardless of initial conditions.  To the extent that that has not happened in the examples shown in Figure~\ref{fig:AllFlows020}, it appears that this equilibration process does not exhibit that aspect of thermalization. There appear to be interesting parallels between the behaviors of the ACL model and the behaviors associated with localization phenomena in condensed matter systems~\cite{Nandkishoredoi:10.1146/annurev-conmatphys-031214-014726}. I will touch on this a couple of times in this paper but warn the reader that I use the term ``thermalize'' with a grain of salt since no example given here has all the features one associates with full thermalization.
 
\section{Varying the Coupling Strength \boldmath{$E_I$}}
\label{sec:varyEI}

It is instructive to consider the special case of $E_I=0$. In that case, no energy will flow between the environment and the SHO, and the energies $\left<H_e\right>$ and $\left<H_s\right>$ are separately conserved. When $E_I=0$, formally, \emph{any} initial state is already ``equilibrated'' in terms of the values $\left<H_e\right>$ and $\left<H_s\right>$, insofar as after an extended period of evolution these will be unchanged. In this section, we examine the different behaviors that emerge as $E_I$ is varied.  We will see that the $E_I=0$ case offers a useful reference point for this exploration and helps us interpret the lack of ``thermalization'' discussed with Figure~\ref{fig:AllFlows020}.

\begin{figure}[h]
\includegraphics[width=3.7 in]{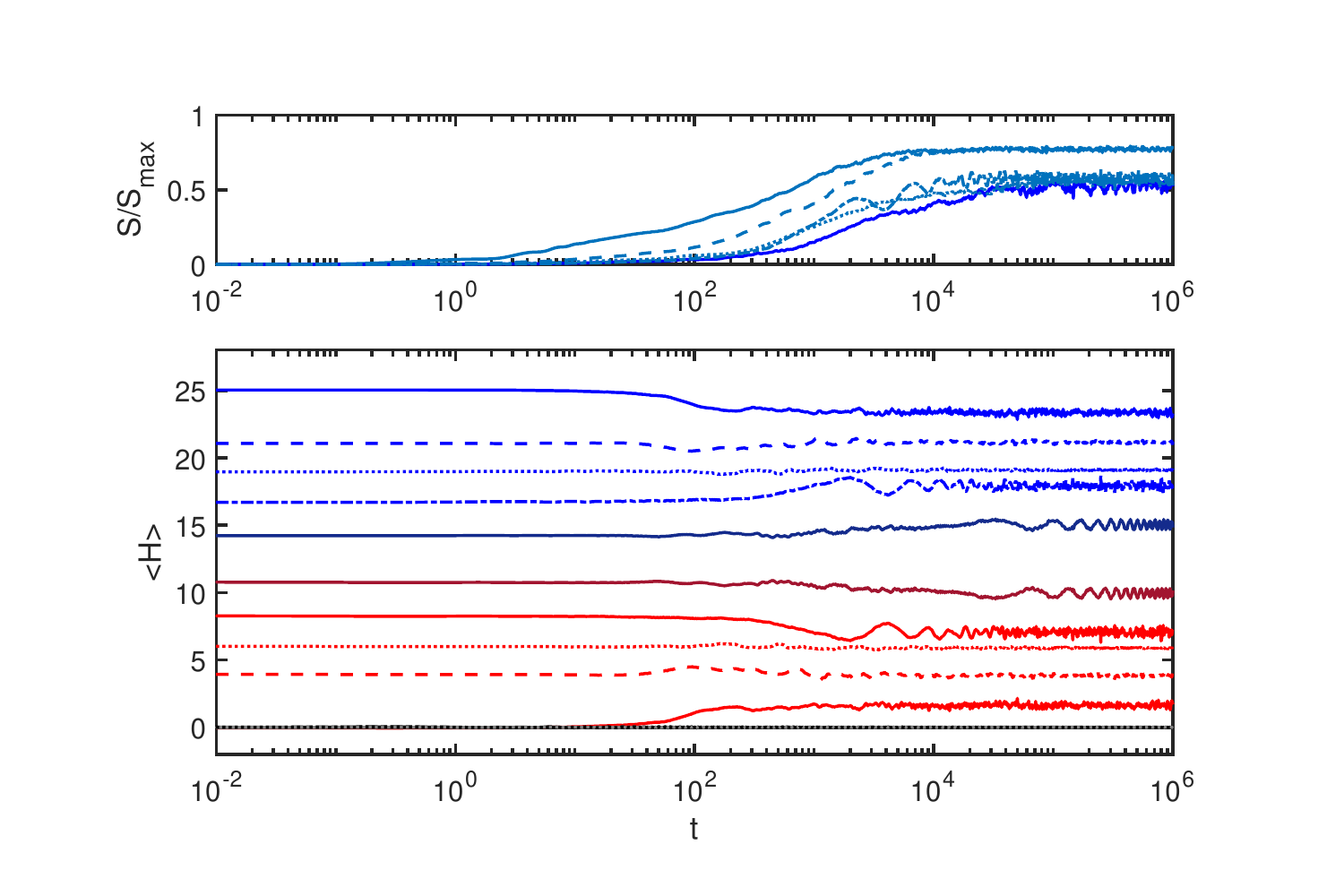}
\caption{{$E_I=0.007$: } 
These curves correspond to the curves in Figure~\ref{fig:AllFlows020}, but evolved with a lower value of $E_I$, closer to the limit of complete decoupling. These curves exhibit less energy flow, and greater differences among the equilibrium values for each subsystem, despite the fixed value of the global energy. This is what one expects as one approaches the $E_I=0$ limit. 
\label{fig:AllFlows007}}
\end{figure}

Figure~\ref{fig:AllFlows007} uses $E_I=0.007$, a factor of $0.35$ down from the case shown in Figure~\ref{fig:AllFlows020}.  The sets of initial conditions and the other parameters in $H_w$ are identical for the two figures.  Here, the energy in each subsystem changes very little, and the equilibrium energy values achieved from different initial conditions are further apart than in Figure~\ref{fig:AllFlows020}.  This is as expected since one is closer to the $E_I=0$ case. {One might have} the intuition that any nonzero value of $E_I$ should allow equilibration, with smaller $E_I$'s leading to longer equilibration times. I believe that intuition is only valid for much larger systems, and, in any case, it is certainly not valid for the results reported here. 

\begin{figure}[h]
\includegraphics[width=3.7 in]{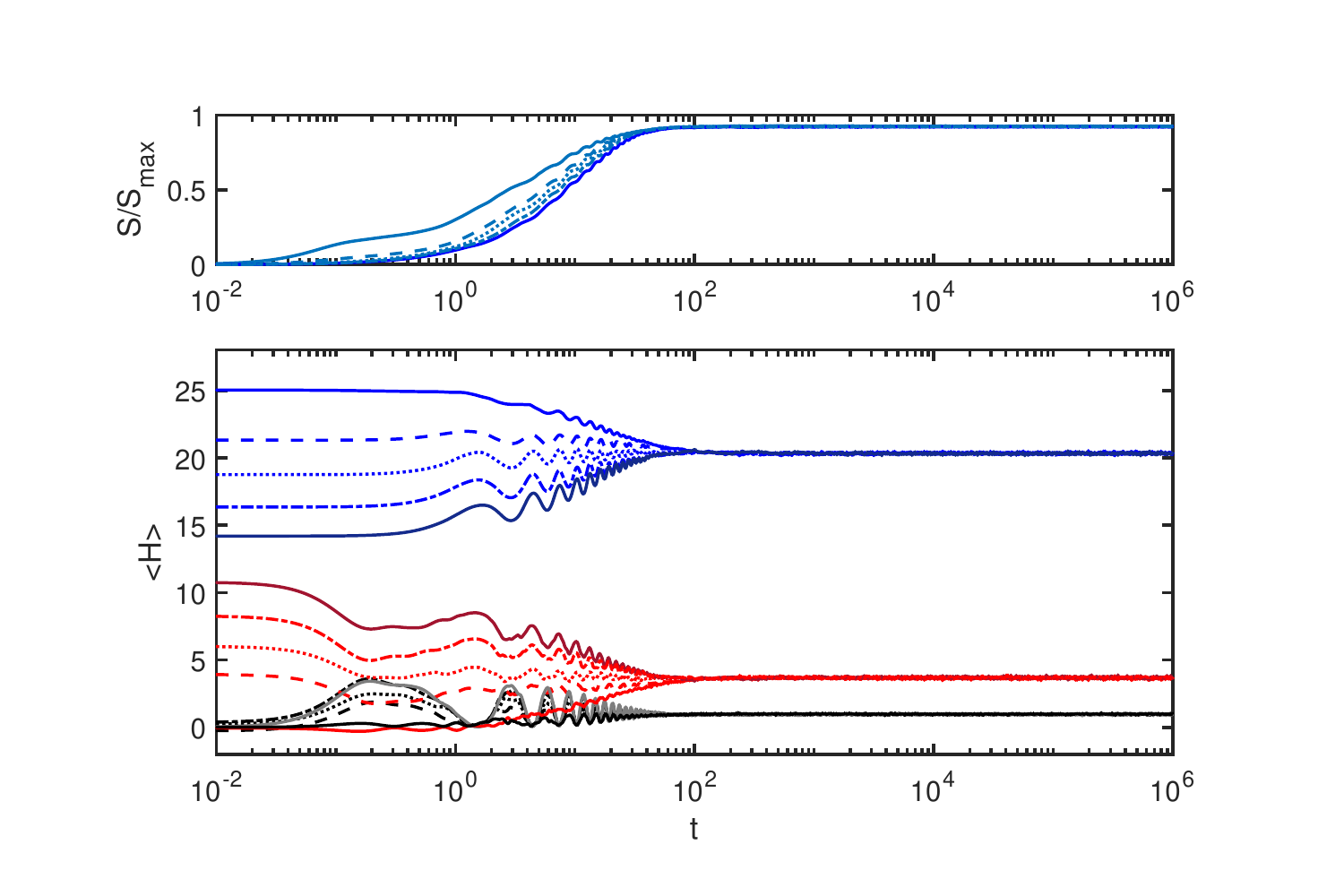}
\caption{{$E_I=0.1$:} 
 Similar to Figures~\ref{fig:AllFlows020} and~\ref{fig:AllFlows007}, but evolved with a larger value of $E_I$ ($5$ times larger than for Figure~\ref{fig:AllFlows020}). All the initial conditions converge to the same energy value for each subsystem, realizing the primitive notion of ``thermalization'' discussed in Section~\ref{sec:EnoT}.  
\label{fig:AllFlows100}} 
\end{figure}

{Figure~\ref{fig:AllFlows100} corresponds to $E_I=0.1$, considerably larger than values used for Figures~\ref{fig:AllFlows020} and~\ref{fig:AllFlows007}}. 
The energy for each subsystem converges to the same value after equilibration, realizing the sense of ``thermalization'' considered in Section~\ref{sec:EnoT}.  I will further explore the ways this equilibration process is different from what is seen for other values of $E_I$ in later sections of this article.  In addition, note that the interaction energy $\left<H_I\right>$ (black curves) shows larger fluctuations than for the previous plots, but that these curves still settle down to the same value, which is considerably smaller than the equilibrium energies in either subsystem. This allows one to still consider this a ``weakly coupled'' case.

\begin{figure}[h]
\includegraphics[width=3.7 in]{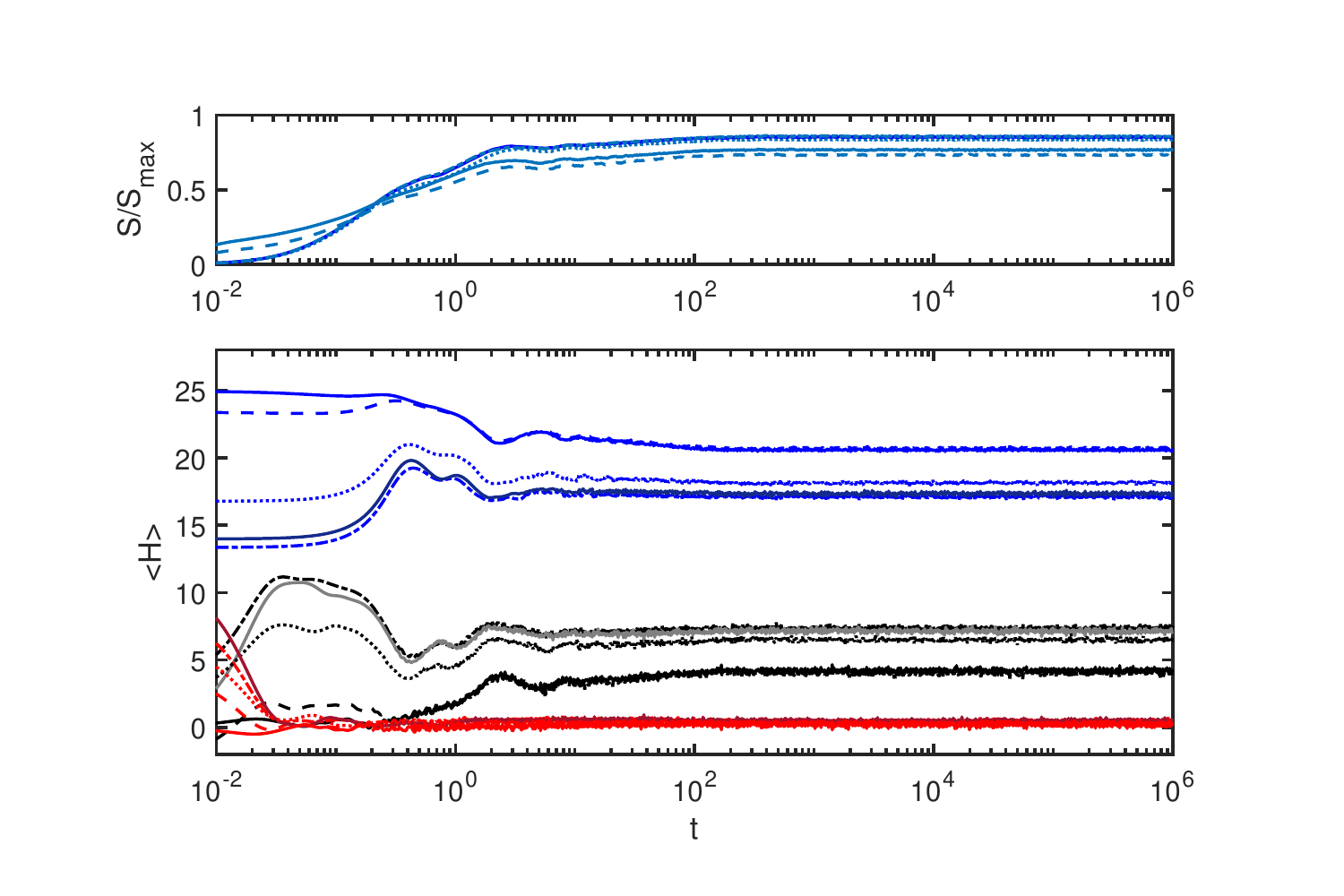}
\caption{{$E_I=1$: This} 
 case is strongly coupled.  Under such conditions, each subsystem has less of an individual identity, but the dephasing process still creates some notion of equilibration. 
\label{fig:AllFlows1}}
\end{figure}

Figure~\ref{fig:AllFlows1} shows a strongly coupled case, with $E_I$ $10$ times larger than the value used for Figure~\ref{fig:AllFlows100}.  
 With such a large coupling, each subsystem has much less of an individual identity in terms of its evolution and the interpretation of $\left<H_s\right>$ and $\left<H_e\right>$.  However, the dephasing process still leads to stable values for $\left<H_s\right>$, $\left<H_e\right>$ and $S$ at later times, with only small fluctuations. 

\section{Energy Distributions}
\label{sec:edist}

Throughout this paper, I take the global system to be in a pure quantum state $\left|\psi_w\right>$.  In general, each subsystem will be described by a density matrix according to
\begin{equation}
   {\rho _s} \equiv T{r_e}\left( {{{\left| \psi  \right\rangle }_w}{}_w\left\langle \psi  \right|} \right)
   \label{eqn:RhoSdef}
\end{equation}
and
\begin{equation}
   {\rho _e} \equiv T{r_s}\left( {{{\left| \psi  \right\rangle }_w}{}_w\left\langle \psi  \right|} \right).
   \label{eqn:RhoEdef}
\end{equation}
When written on the basis of eigenstates of $H_s$ and $H_e$, respectively, the diagonal elements of $\rho_s$ and $\rho_e$ give the probabilities assigned to different values of the subsystem energies. I define 
\begin{equation}
    P_s(E) \equiv diag(\rho_s^E)
        \label{eqn:Psdef}
\end{equation}
and
\begin{equation}
    P_e(E) \equiv diag(\rho_e^E)
        \label{eqn:Pedef}   
\end{equation}
where the superscript $E$ indicates that the energy eigenbasis is used. Note that, for the ACL model, the argument $E$ on the left side of these expressions is drawn from the discrete set of energy eigenvalues. 
These figures show the case for which the energies (namely, the first moments of these distributions) are shown in Figure~\ref{fig:FirstFlow}.  One can see that, as we have already noted in the case of the first moments, the whole distribution stabilizes at late times, up to small fluctuations. 

\begin{figure}[h]
\includegraphics[width=3.7 in]{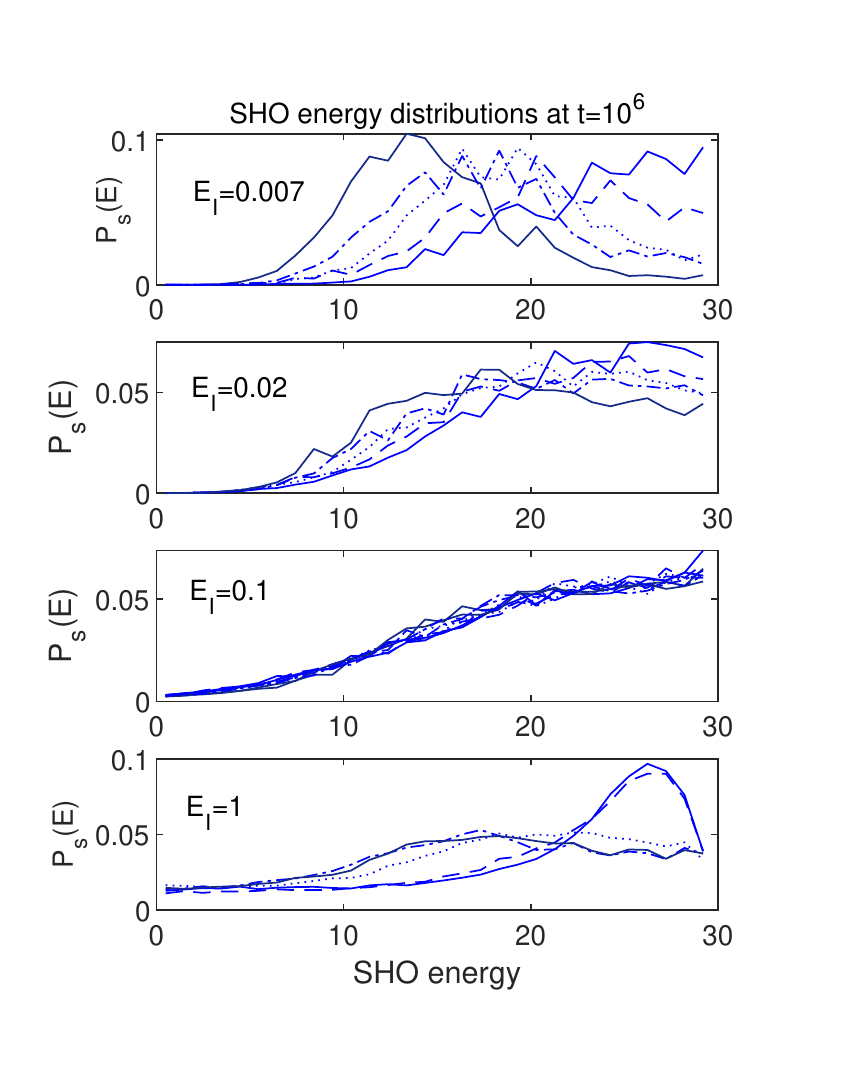}
\caption{{Late time energy distributions} 
 for the SHO.  Each panel shows a set corresponding to the five different initial conditions, evolved using the value of $E_I$ as marked. For $E_I=0.1$, the SHO energies (first moments of $P_s(E)$) converged at late times for the different initial conditions (see Figure~\ref{fig:AllFlows100}). These results indicate that many more than just the first moment converges for this value of $E_I$. The $E_I=0.1$ panel includes the phase randomized version of each curve as well. 
\label{fig:NisLateSetMultiEi}}
\end{figure}

\begin{figure}[h]
\includegraphics[width=3.7 in]{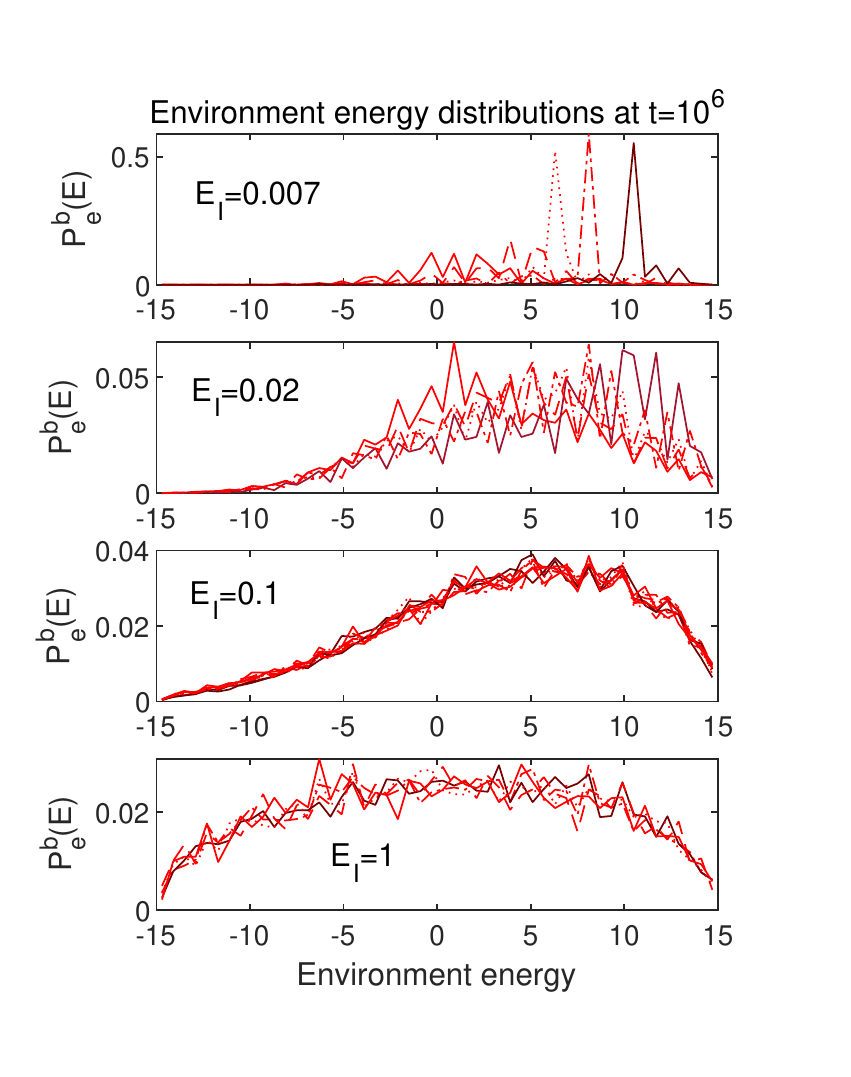}
\caption{{Late time energy} 
 distributions for the environment. Each panel shows a set corresponding to the five different initial conditions, evolved using the value of $E_I$ as marked. For $E_I=0.1$, the environment energies (first moments of $P_e(E)$) converged at late times for the different initial conditions (see Figure~\ref{fig:AllFlows100}). These results indicate that many more than just the first moment converges for this value of $E_I$. The $E_I=0.1$ panel includes the phase randomized version of each curve as well. 
\label{fig:NieLateSetMultiEi}}
\end{figure}

Figures~\ref{fig:NisLateSetMultiEi} and~\ref{fig:NieLateSetMultiEi} show the late time distributions for the full range of initial conditions and choices of $E_I$ considered above. It is certainly not surprising that the cases where the subsystem energies equilibrated to different values show significantly different late time forms for the overall distributions.  It is interesting though, that in the case ($E_I=0.1$) where the subsystem energies equilibrated to the same values for different initial states the entire distribution appears to equilibrate to the same form, encompassing many more moments than just the first.  

\begin{figure}[h]
\includegraphics[width=3.7 in]{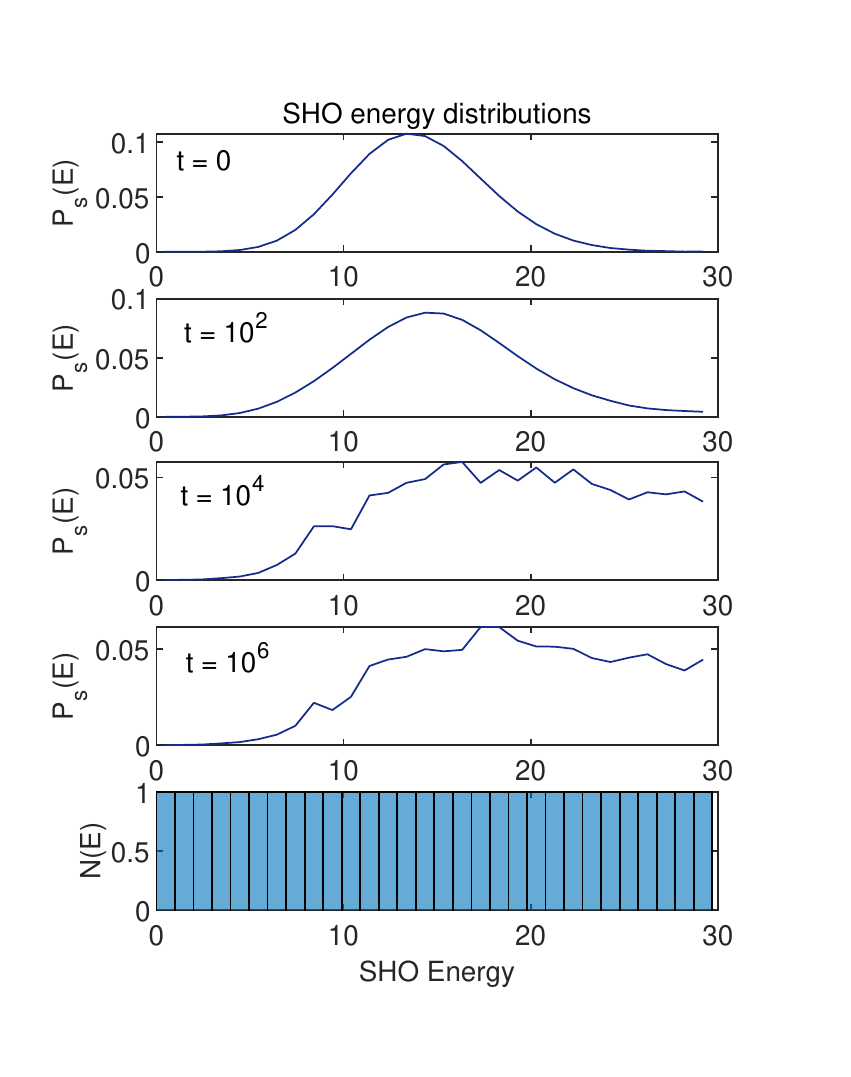}
\caption{Time evolution of SHO energy distribution $P_s(E)$ for the case shown in Figure~\ref{fig:FirstFlow}.  The histogram in the lower panel is the density of energy eigenstates for the SHO (which is uniform). I have connected the discrete set of points given by $P_s(E)$ here for ease of viewing. 
\label{fig:NisOFtFirst}}
\end{figure}

\begin{figure}[h]
\includegraphics[width=3.7 in]{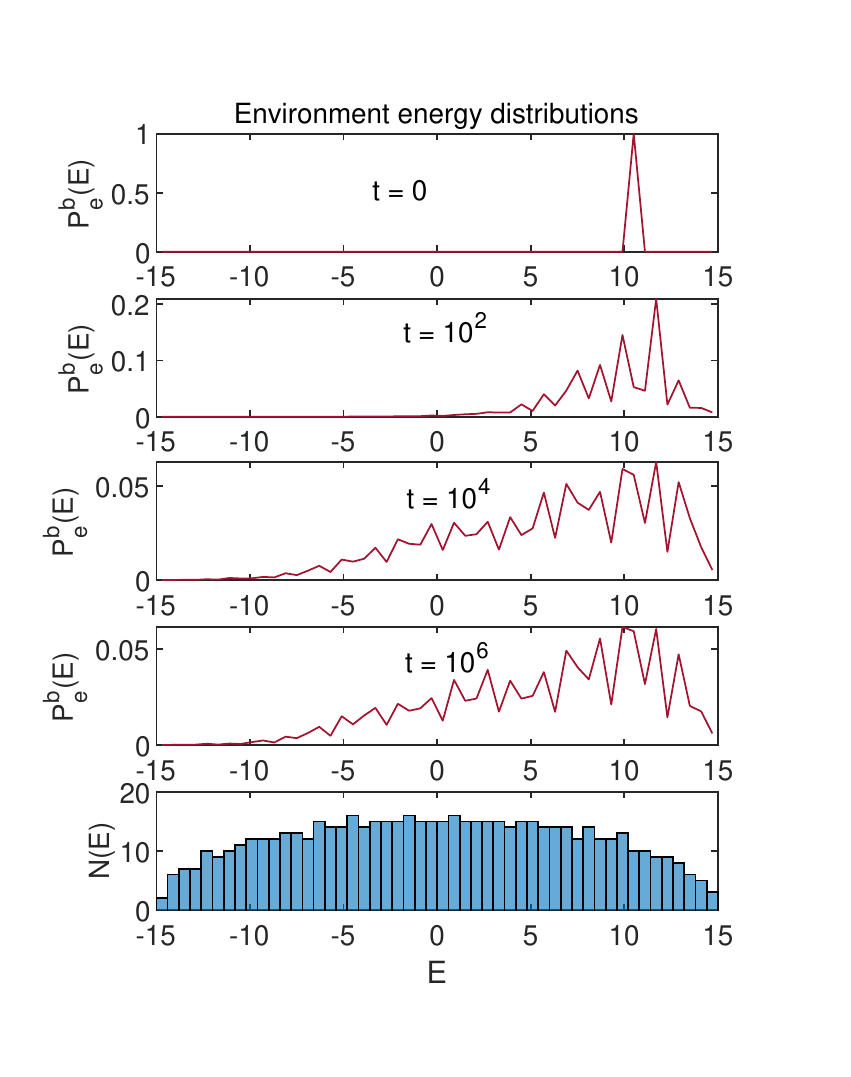}
\caption{{Time evolution of} 
 environment energy distribution $P_e(E)$ for the case shown in Figure~\ref{fig:FirstFlow}. The histogram in the lower panel shows the density of energy eigenstates for the environment (which reflects the Wigner semicircle form expected for a random Hamiltonian).  The $N_e = 600$ different eigenvalues have been binned as shown in the histogram and I plot $P_e^b(E)$, which is the total probability in the corresponding bin.  This discrete set of points is connected for visualization purposes.  
\label{fig:NieOFtFirst}}
\end{figure}

Figure~\ref{fig:NisOFtFirst} shows the time evolution of $P_s(E)$ and Figure~\ref{fig:NieOFtFirst} shows the corresponding $P_e(E)$ evolution.

In Section~\ref{sec:EnoT}, I discussed a primitive notion of ``thermalization'' based on the expectation that a thermalized system should share energies in the same proportions among different subsystems, regardless of the initial state, as long as each initial state had the same total energy.  Among the cases considered in Section~\ref{sec:varyEI}, we saw that only the $E_I=0.1$ case met that criterion. Figures~\ref{fig:NisLateSetMultiEi} and~\ref{fig:NieLateSetMultiEi} show that, for $E_I=0.1$, the system meets a stronger criterion, namely that many moments of the final energy distributions are independent of the initial state. This is certainly what one gets in the case of true thermalization of realistic physical systems, although it is worth emphasizing that none of the distributions shown are truly thermal in the sense of having the Gibbs form, as a function of an actual temperature. Still, having a parameter to dial which can turn on or off the rudimentary features of thermalization discussed here suggests that further explorations might reveal some insights into the notion of thermalization in general.  I undertake such explorations in what follows. 

{It is tempting} to make contact with the notion of ``generalized canonical state'' as discussed, for example, in~\cite{Popescu2006NatPh...2..754P}. However, in the places I have seen the generalized Gibbs distribution discussed it has taken a more idealized form.  For example, a thermodynamic limit is taken or an idealized notion of ``passivity''~\cite{RigolPhysRevLett.98.050405,lenard1978thermodynamical} is utilized. Those idealizations would preclude the sort of small fluctuations that appear in the my results. 

Before concluding this section, I want to comment about the strongly coupled $E_I=1$ case. I have included it in this paper for completeness, but it should be emphasized that, due to the strong coupling, there is little meaning to the $s$ and $e$ subsystems. The quantities plotted in that case ($\left<H_s\right>$, $\left<H_s\right>$, $P_s(E)$ and $P_e(E)$), are mathematically well-defined, but they do not have natural physical interpretations.  It seems unlikely that the $E_I=1$ case admits a physically useful interpretation of the full space as a tensor product of any subspaces---certainly not the specific $e$ and $s$ ones considered here.  One point one can make about this case is that it demonstrates that a rudimentary process of equilibration, driven by dephasing, is possible without any reference to energy flow, or any sense in which one subsystem is acting as a ``bath'' to another. Comments along these lines appear in~\cite{Popescu2006NatPh...2..754P}. 

\section{The \boldmath{$H_w$} Eigenstates}
\label{sec:w_eigenstates}

\subsection{Energy Distributions in the Subspaces}
\label{distin_s_e}
Here, I explore the relationship between the phenomena discussed above and the form of the eigenstates of the global Hamiltonian $H_w$.  I will focus here on the $E_I=0.007$ and $E_I=0.1$ cases. The $E_I= 0.02$ case exhibits behavior intermediate between those two, and, as discussed above, the $E_I=1$ is not amenable to deeper analysis due to the \mbox{strong coupling. }

Recall that, for the $E_I=0$ case, the eigenstates of $H_w$ are products of eigenstates of $H_s$ and $H_e$. Once the interaction is turned on, in general, $H_w$ eigenstates will appear as density matrices in the $s$ and $e$ subspaces.  I will utilize the techniques from Section~\ref{sec:edist} to focus on the energy distributions $P_s(E)$ and $P_e(E)$, given by the diagonal elements of the density matrices according to Equations~\eqref{eqn:Psdef} and~\eqref{eqn:Pedef}.  While these give incomplete information (only certain matrix elements will be shown, and $P_e(E)$ will appear binned as above), that information is sufficient to get a sense of what is going on.  I have studied the properties of these states with more complete information than presented here and have confirmed that the information I do present gives a reasonable characterization for the points I want \mbox{to make.} 

\begin{figure}[h]
\includegraphics[width=3.7 in]{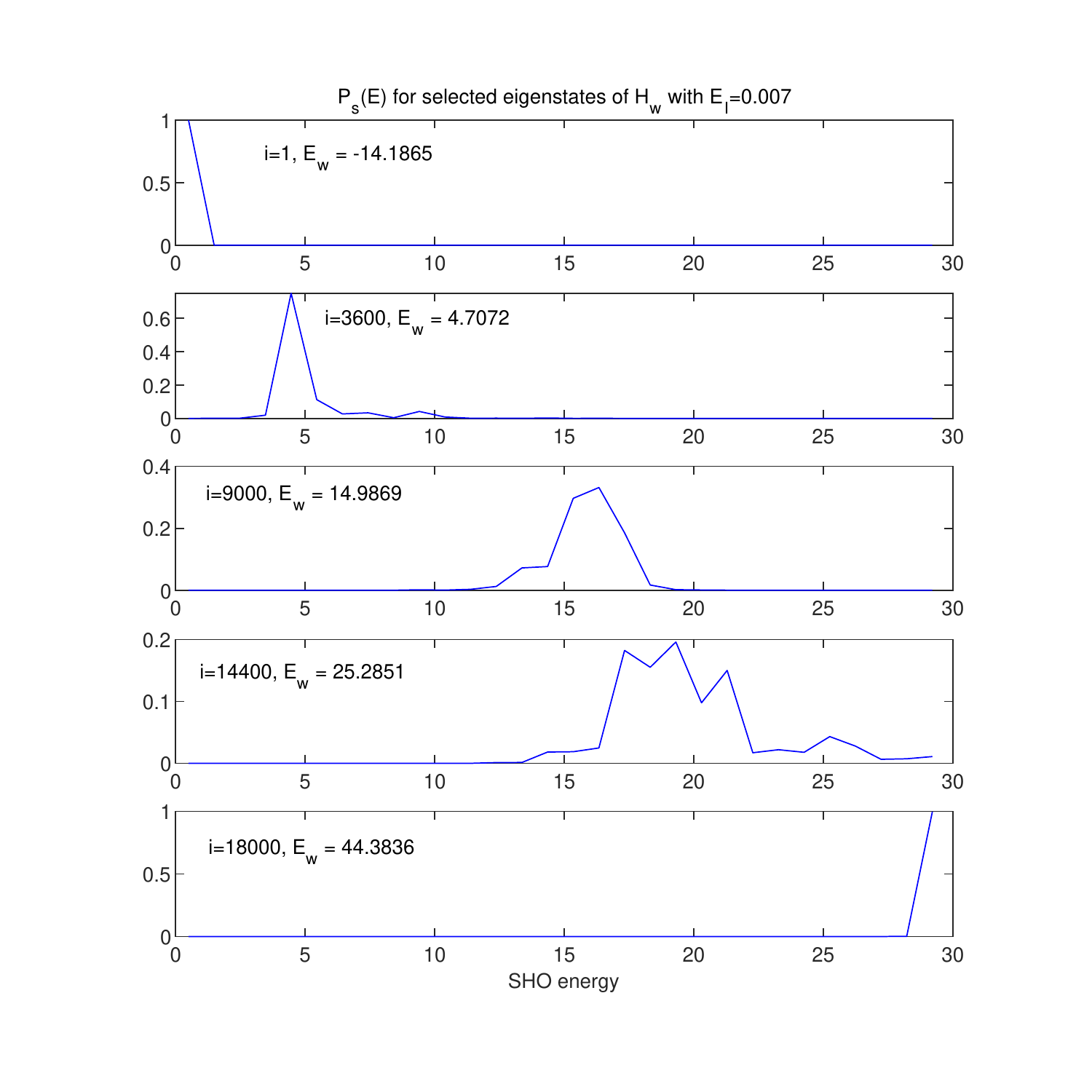}
\caption{{A selection of eigenstates} 
 of $H_w$ (running from minimum to maximum eigenvalues) represented in terms of distributions in SHO energy. Here, $E_I = 0.007$. Note that, while these are not perfect delta functions, they are reasonably sharply peaked, as one would expect for a situation close to the $E_I = 0$ limit. 
\label{fig:Eigs_007_broad_s}}
\end{figure}

\begin{figure}[h]
\includegraphics[width=3.7 in]{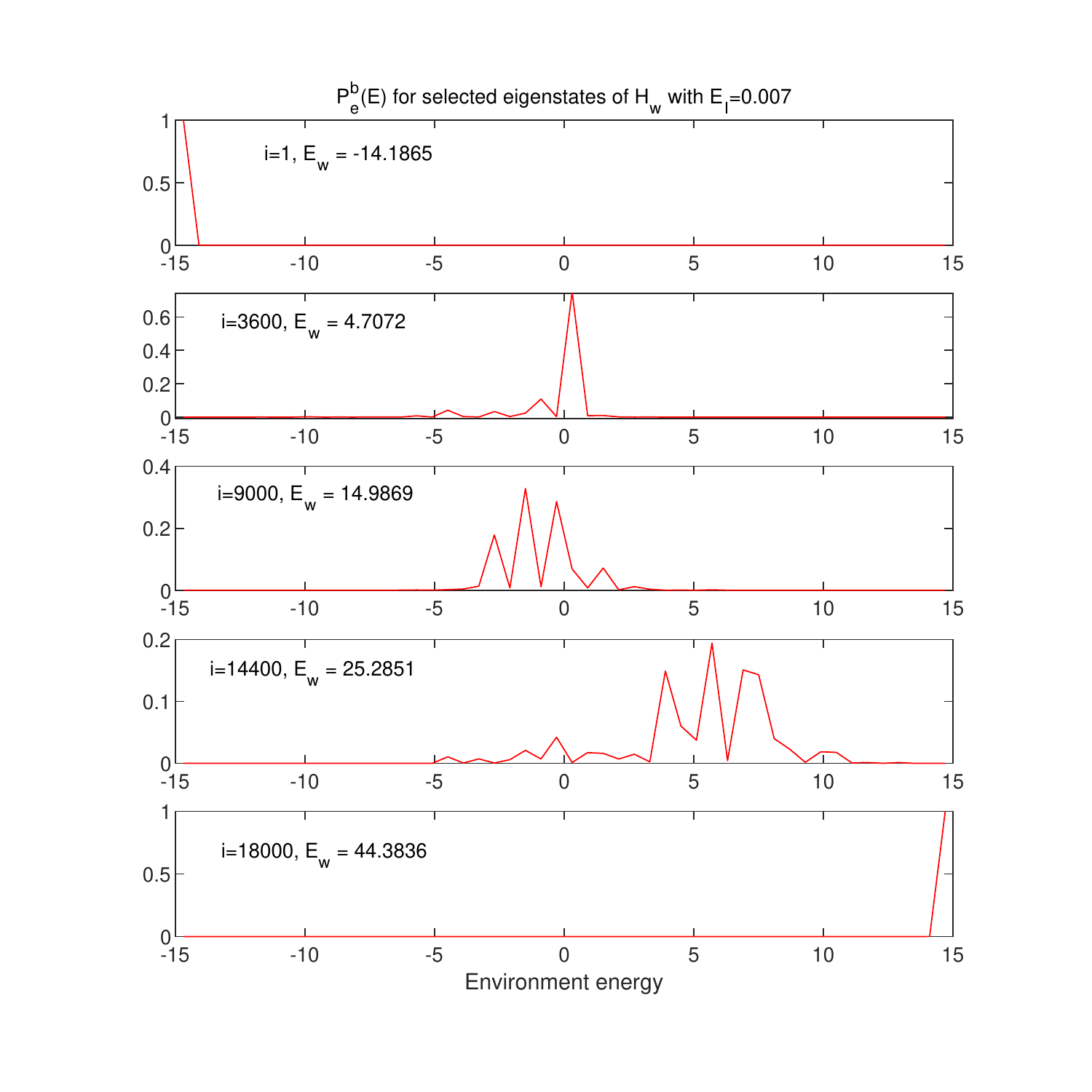}
\caption{{The same eigenstates}   
 shown in Figure~\ref{fig:Eigs_007_broad_s} are shown here in terms of the (binned) distributions in environment energy.  These too are reasonably sharply peaked, as expected. 
\label{fig:Eigs_007_broad_e}}
\end{figure}

Figures~\ref{fig:Eigs_007_broad_s} and~\ref{fig:Eigs_007_broad_e} show information about a broad range of $H_w$ eigenstates in terms of $P_s(E)$ and $P_e(E)$, respectively, for the $E_I=0.007$ case. 
 For this case, one can see that, while the eigenstates of $H_w$ may not be perfect product states of $H_s$ and $H_e$ eigenstates, the energy distributions are still quite localized, as one would expect for very weak coupling. 

\begin{figure}[h]
\includegraphics[width=3.7 in]{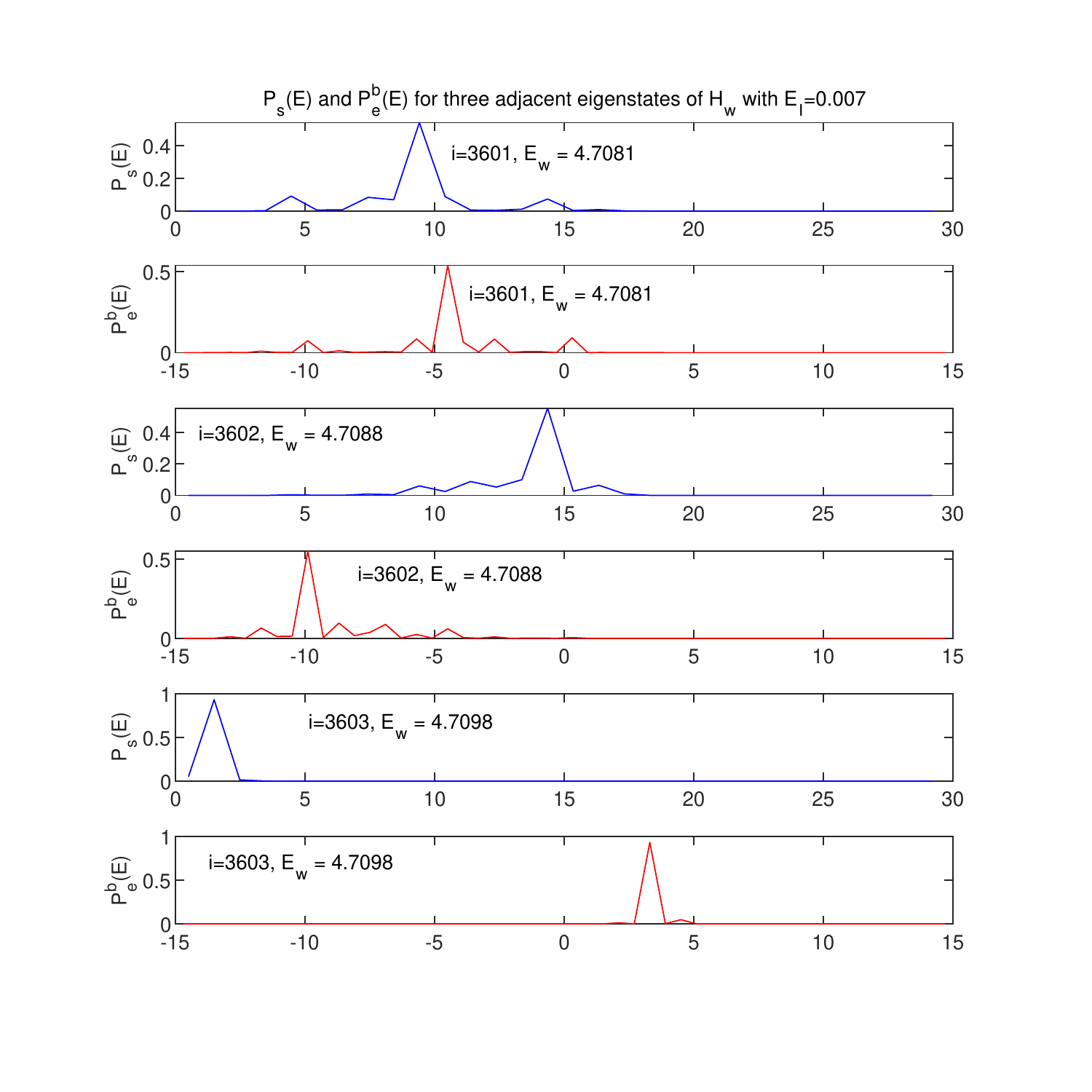}
\caption{{Energy distributions}  
 shown in both $s$ (blue) and $e$ (red) for three adjacent eigenstates of $H_w$, with $E_I=0.007$. Although the associated values of $E_w$ 
are essentially identical the energy is distributed in very different ways between the two subsystems.
\label{fig:Eigs_007_close_1}}
\end{figure}

Figure~\ref{fig:Eigs_007_close_1} shows energy distributions in both $s$ and $e$ subspaces for three adjacent energy eigenstates of $H_w$.
 While the three pairs of curves represent eigenvalues of $H_w$ which differ by at most $0.04\%$, the energy distribution between $s$ and $e$ is very different for each of the three cases. This is simply a reflection of the fact that there \emph{are} many ways of distributing a fixed total energy among the two subsystems, and the peaked nature of the energy distributions for this very weakly coupled case allows that fact to play out in a simple and vivid way in the eigenstates of $H_w$. 
\unskip

\begin{figure}[h]
\includegraphics[width=3.7 in]{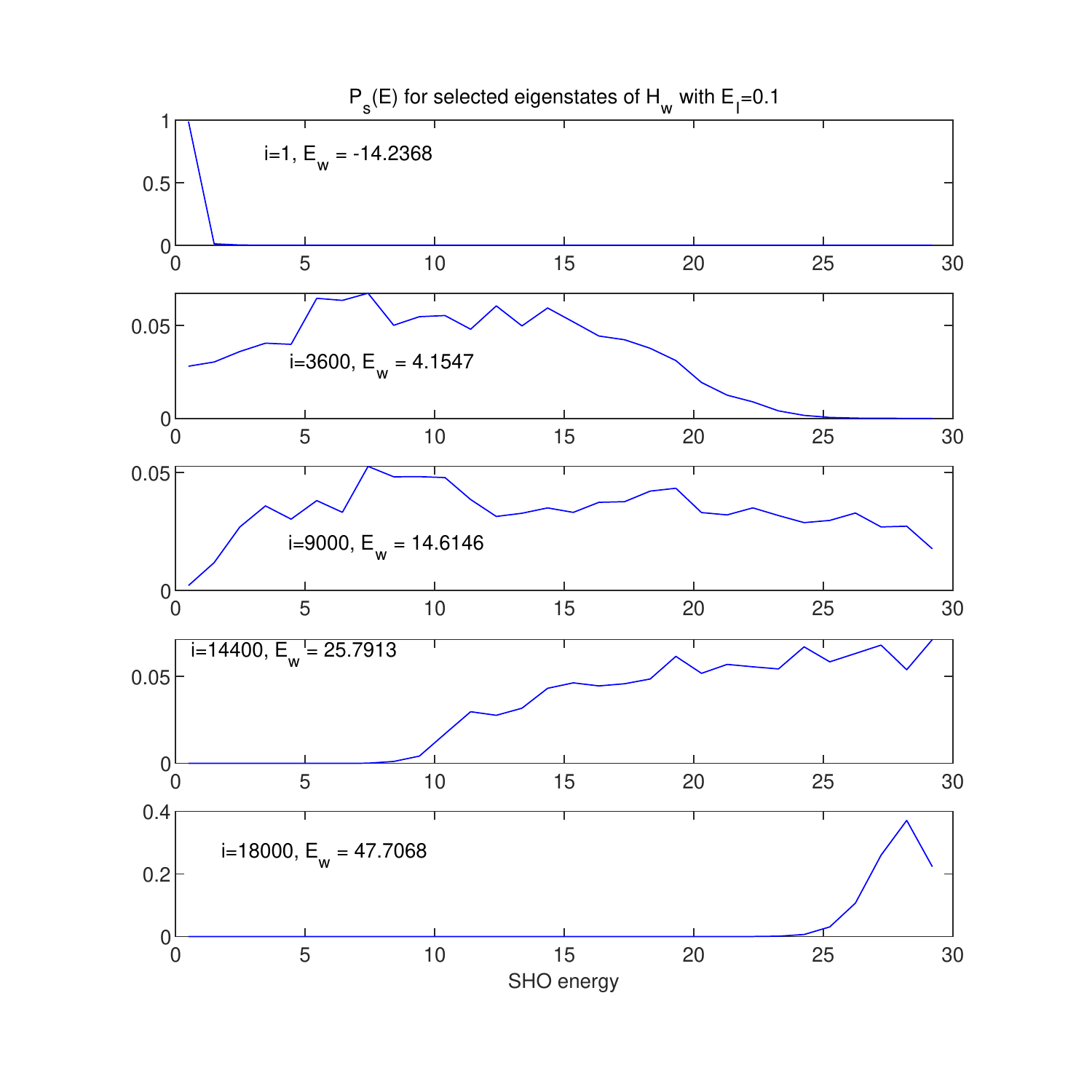}
\caption{{A selection} 
 of eigenstates of $H_w$ (running from minimum to maximum eigenvalues) represented in terms of distributions in SHO energy. Here, = $E_I = 0.1$. While this case is weakly coupled by some measures, the interaction is strong enough to mix many of the energy eigenstates of the SHO, creating much broader distributions than seen in Figure~\ref{fig:Eigs_007_broad_s} for the $E_I=0.007$ case, at least away from the extreme ends of the spectrum. 
\label{fig:Eigs_100_broad_s}}
\end{figure}

\begin{figure}[h]
\includegraphics[width=3.7 in]{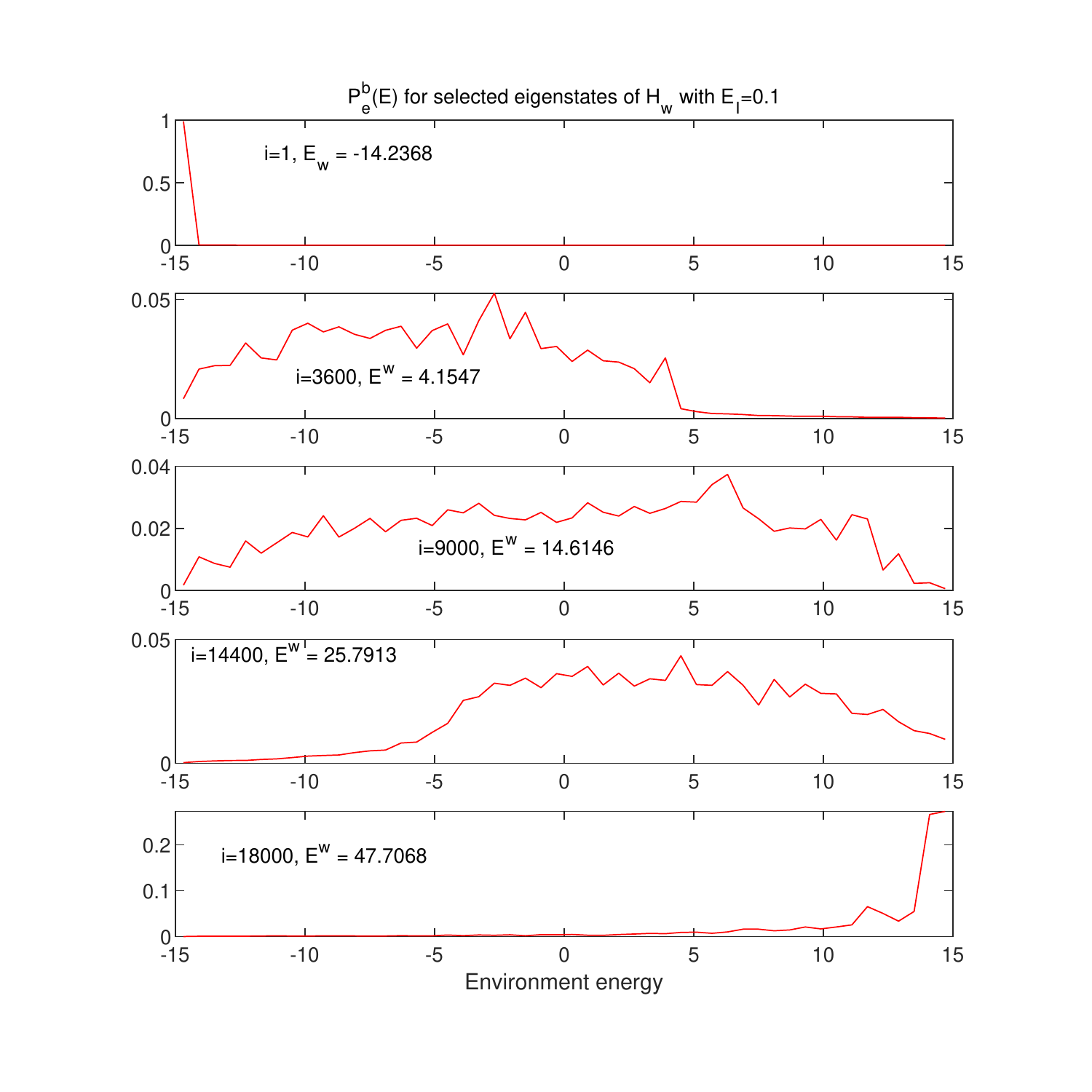}
\caption{{The same eigenstates} 
 shown in Figure~\ref{fig:Eigs_100_broad_s} are shown here in terms of the distributions in environment energy.  These distributions are also much more broad than those shown in Figure~\ref{fig:Eigs_007_broad_e} for the $E_I=0.07$ case. 
\label{fig:Eigs_100_broad_e}}
\end{figure}

Figures~\ref{fig:Eigs_100_broad_s} and~\ref{fig:Eigs_100_broad_e} show information about a broad range of $H_w$ eigenstates in terms of $P_s(E)$ and $P_e(E)$, respectively, this time for the $E_I=0.1$ case. 
While in many respects the $E_I= 0.1$ case might be thought of as ``weakly coupled''---for example, note the small relative values of the interaction energy shown in black in Figure~\ref{fig:AllFlows100}---the interaction term is strong enough to mix many eigenstates of $H_s$ and of $H_e$, leading to much broader distributions, except at the extremes of the spectrum.  Note that the relevant measure for understanding the breadth of these distributions is the size of the interaction term relative to the \emph{spacing} of the energy eigenvalues of the respective subsystems. 

\begin{figure}[h]
\includegraphics[width=3.7 in]{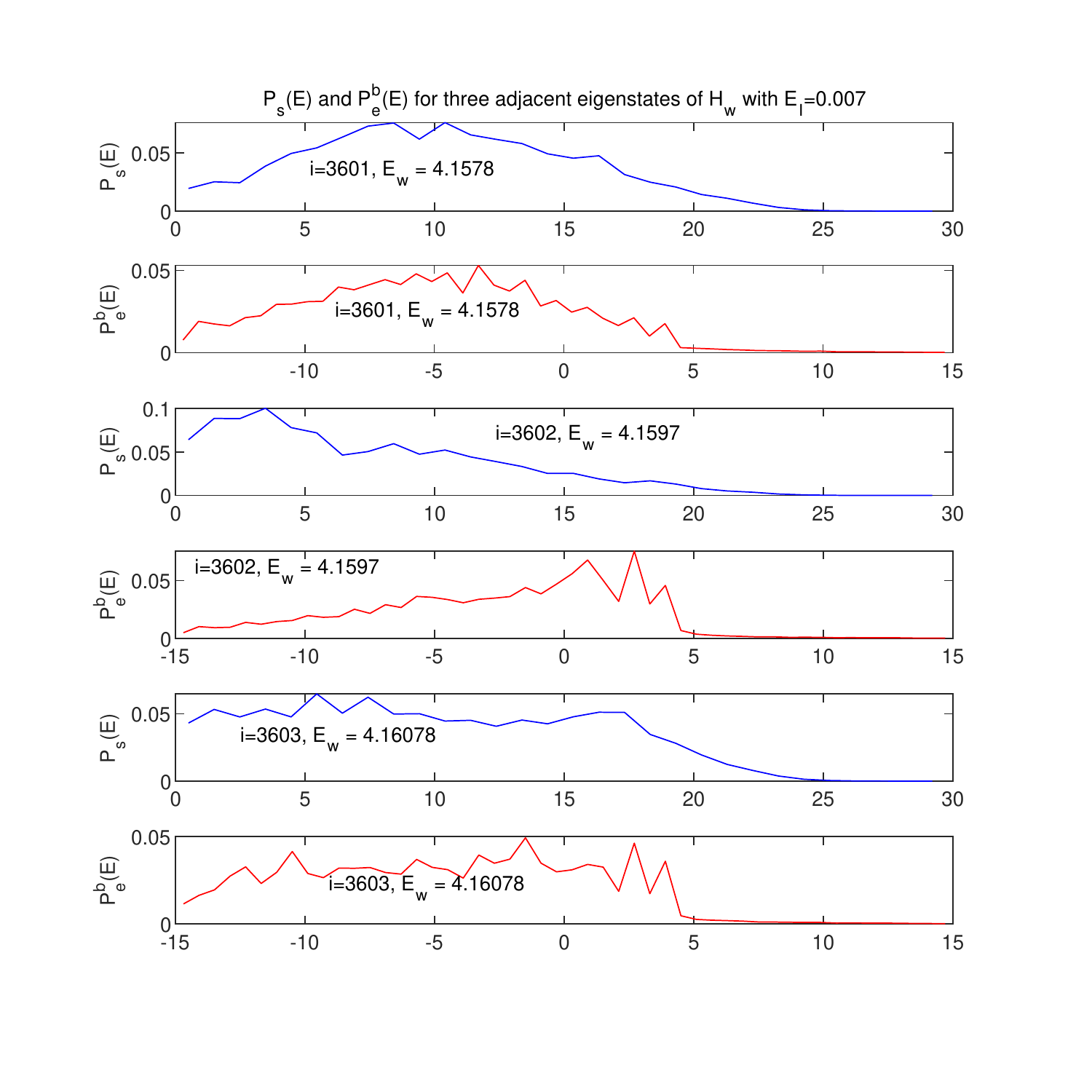}
\caption{{Energy distributions shown}   
 in both $s$ (blue) and $e$ (red) for three adjacent eigenstates of $H_w$ with $E_I=0.1$. For these broad distributions, the differences between neighboring eigenstates are more subtly compared with the more weakly coupled case shown in Figure~\ref{fig:Eigs_007_close_1}.
\label{fig:Eigs_100_close_1}}
\end{figure}

Figure~\ref{fig:Eigs_100_close_1} shows energy distributions in both $s$ and $e$ subspaces for three adjacent energy eigenstates of $H_w$, here with $E_I = 0.1$.
 In contrast with what was seen for the more weakly coupled case in Figure~\ref{fig:Eigs_007_close_1}, the changes in the subsystem energy distributions as one steps between adjacent eigenstates of $H_w$ are more subtle, although the differences can be discerned upon inspection. 

\subsection{Energy Distributions in the Global Space $w$}
\label{distin_w}

\begin{figure}[h]
\includegraphics[width=3.7 in]{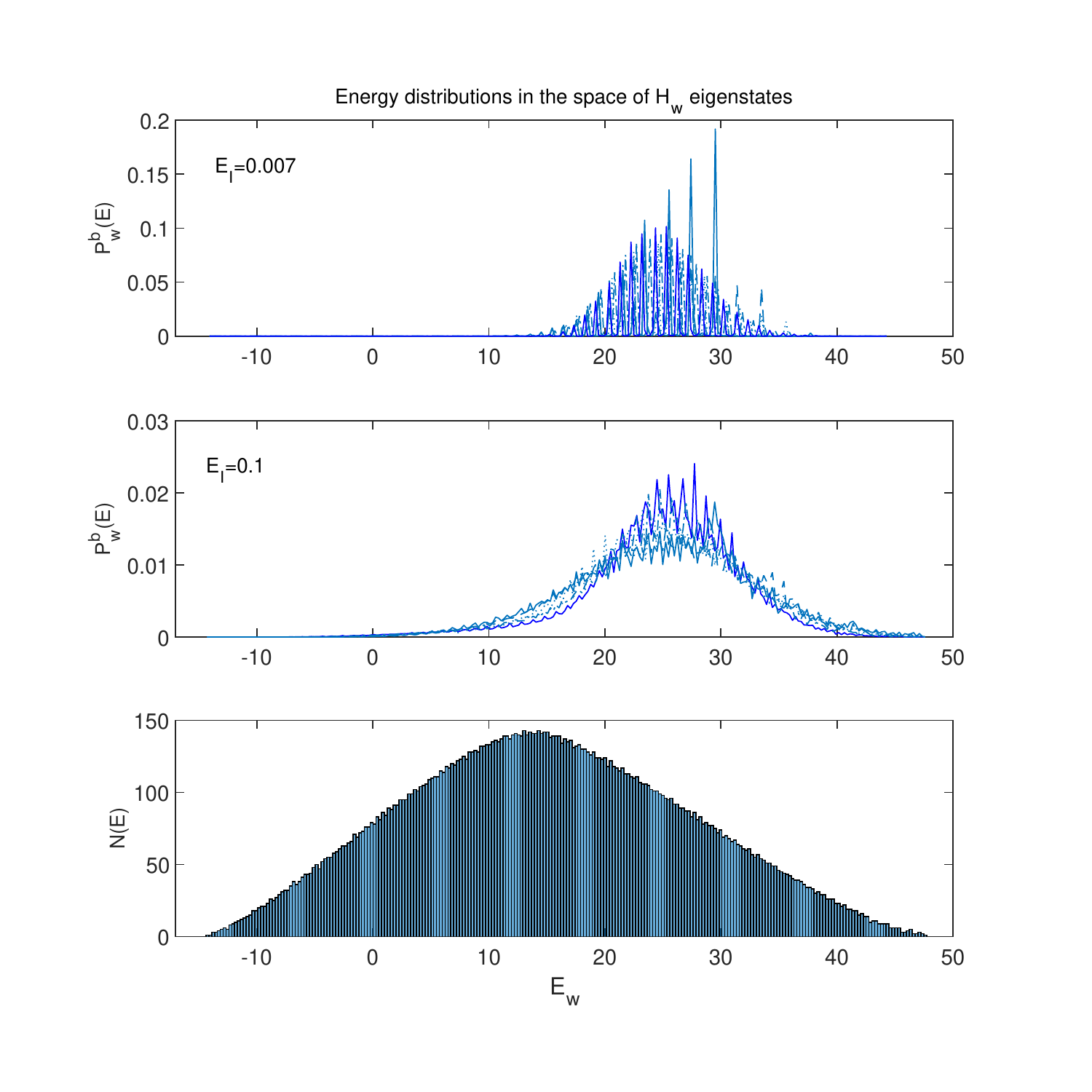}
\caption{{The (binned) global}  
 energy distributions $P_w^b(E)$ corresponding to the five initial states used throughout this paper. One can see that these curves have very different behaviors depending on the coupling strength $E_I$. For example, the $E_I=0.007$ curves frequently approach zero, and the $E_I= 0.1$ curves do not. The bottom panel shows a histogram representing the density of energy eigenstates $N(E_w)$.  Technically, $N(E_w)$ will be different for the two values of $E_I$, but both the values shown here are small enough not to change the form of $N(E_w)$ significantly. 
\label{fig:P34forPaper}}
\end{figure}

\begin{figure}[h]
\includegraphics[width=3.7 in]{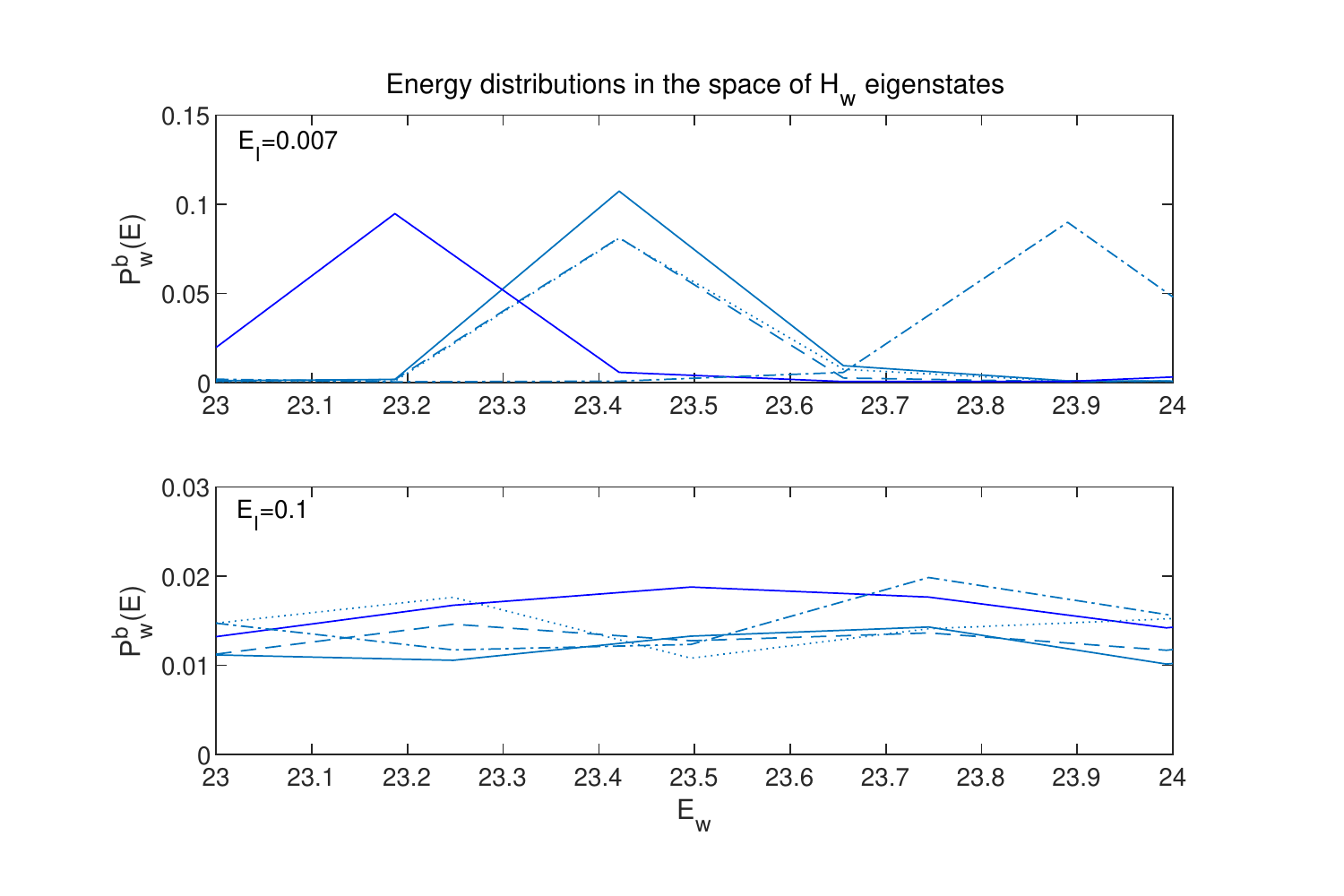}
\caption{A zoomed-in look at the first two panels of Figure~\ref{fig:P34forPaper}. One can see that not only do the $E_I=0.007$ curves approach zero frequently, in contrast to the $E_I=0.1$ curves, but the curves for each state exhibit very different patterns of large and small values. 
\label{fig:P34forPaperZ}}
\end{figure}

Having established significant differences between eigenstates of $H_w$ as they appear in the subsystems, depending on the strength of the coupling $E_I$, I will now consider how the initial conditions are represented in these different sets of eigenstates. Figure~\ref{fig:P34forPaper} shows the distributions $P_w(E)$, defined in the same manner as $P_s$ and $P_e$ (Equations~\eqref{eqn:Psdef} and~\eqref{eqn:Pedef}) but in the global ``world'' space $w$ using $\rho_w = {\left| \psi  \right\rangle }_w{}_w\left\langle \psi  \right|$.  The $P_w$'s are time independent, and the different curves correspond to the set of five different initial conditions used to create each of Figures~\ref{fig:AllFlows020}--\ref{fig:AllFlows1}. (Note that I have only been using the same set of five initial conditions throughout this paper.  I have been evolving and analyzing each one using $H_w$'s with different values of $E_I$.)
  Figure~\ref{fig:P34forPaperZ} shows zoomed-in portions of the top two panels of Figure~\ref{fig:P34forPaper}.
 The large oscillations of the $E_I=0.007$ curves are especially clear in Figure~\ref{fig:P34forPaperZ}, and one can see that the individual curves have very different locations of their peaks and minima. 

\subsection{Interpretation}
\label{sec:InterpAnal}
I have presented information about the eigenstates of $H_w$ using both the subsystem and global perspectives. These perspectives can be brought together in the following way.  Consider the state $\left|E_i\right>_s\left|E_j\right>_e$, a product of eigenstates of $H_s$ and $H_e$ (with particular values of $i$ and $j$), and consider expanding that state in eigenstates of $H_w$. For the $E_I=0.007$ case, if a particular eigenstate $\left|E_k\right>_w$ of $H_w$ has a strong overlap with $\left|E_i\right>_s\left|E_j\right>_e$, then the $(k+1)$th state is likely to have a much weaker overlap. This is expected given the way the energy distributions shift among the subsystems as the index is incremented, as illustrated in Figure~\ref{fig:Eigs_007_close_1}. ({A similar situation} is considered for many body systems in~\cite{IkedaPhysRevE.84.021130}.) 
Expanding in eigenstates of $H_w$ with $E_I=0.1$ will work very differently. As illustrated in Figure~\ref{fig:Eigs_100_close_1}, neighboring eigenstates will have energy distributions in the subsystems which are not radically different as the index is incremented. This suggests that the overlaps will vary much more smoothly with the index of $E_k$.  

Similarly, the breadth of the distributions of $\left|E_w\right>$'s in the $s$ and $e$ energy distributions for $E_I=0.1$ suggests initial states with energy shared differently between $s$ and $e$ can still pick up similar overlaps with the $\left|E_w\right>$'s, accounting for the overall shape similarity among the different curves in the 2nd panel of Figure~\ref{fig:P34forPaper}.  Furthermore, since $P_w(E)$, plus the phases, gives complete information about the global state $\left|\psi\right>_w$, it is not surprising that, under equilibrium conditions (when the phases may be taken as random), states with similar $P_w(E)$'s also give similar $P_s(E)$'s and $P_e(E)$'s.  ({The} $\left|E_w\right>$ energy distributions in the $s$ and $e$ are not perfectly broad, so, not surprisingly, I have found examples of other initial states with particularly extreme energy distributions among $s$ and $e$ which have somewhat different shapes for $P_w(E)$, and even the 2nd panel of Figure~\ref{fig:P34forPaper} shows noticeable variations.) 

The initial states studied here are products of coherent states in $s$ with energy eigenstates in $e$.  That makes the simple illustration above less rigorous, but the coherent states are somewhat localized in energy, so the main thrust of the illustration should carry through. In addition, the energy distributions in $s$ and $e$ only contain some of the information relevant for calculating the overlap $(_s\left<E_i\right|_e\left<E_j\right|)\cdot\left|E_k\right>_w$, but again that information seems enough to capture some sense of what makes the $P_w(E)$ curves so different for the two values of $E_I$.  Similar arguments can be used to relate my results to the Eigenstate Thermalization Hypothesis, which I do in Appendix~\ref{app:ETH}. 

Finally, if one considers some process of extending this analysis to larger systems, one could imagine cases where the $P_w(E)$ distributions become more narrow (perhaps einselected into sharp energies through weakly coupled environments as in the ``quantum limit'' discussed in~\cite{Paz:1998ib}).  The smooth qualities of $P_w(E)$ we see for $E_I=0.1$ could correspond in such a limit to a relatively flat distribution within the allowed range.  This could connect with ergodic ideas which count each state equally within allowed energies, making contact with conventional statistical mechanics.   

On the other hand, looking at the $E_I=0.007$ case suggests another limit where $P_w(E)$ could remain more jagged, preventing simple statistical arguments from taking hold.  By envisioning limits in this way, it does seem like the primitive ``thermalized'' behaviors of the $E_I=0.1$ case discussed in Sections~\ref{sec:varyEI} and~\ref{sec:edist} are in some sense precursors to a full notion of thermalization for larger systems. Likewise, the alternative limit suggested by the $E_I=0.007$ case has parallels with Anderson and many body localization in large systems.  The localized systems exhibit a lack of thermalization for the same sorts of reasons as the toy model considered here, namely the lack of full access to states that should be allowed based purely on energetic reasons. In addition, just as the localized case appears to reflect additional (approximately) conserved quantities~\cite{SerbynPhysRevLett.111.127201}, I have associated the special features of the $E_I=0.007$ case with the (partially broken) symmetry conserving $\left<H_s\right>$ and $\left<H_e\right>$ separately in the $E_I=0$ limit. 

\subsection{The Effective Dimension as a Diagnostic}
\label{sec:deff}

One way to characterize the different qualities of the sets of curves in Figures~\ref{fig:P34forPaper} and~\ref{fig:P34forPaperZ} is using the ''effective dimension''
\begin{equation}
    d_{eff}^w \equiv \frac{1}{\sum_i (P^w(E_i))^2}.
    \label{eqn:deffdef}
\end{equation}
This quantity takes its minimum value of unity if $P(E_i)$ is a delta function, and reaches its maximum possible value, $N_w$, if all $P(E)$'s are identical.  Table~\ref{tab:deffs} compiles information about the $d_{eff}^w$ values for the curves shown in Figure~\ref{fig:P34forPaper}, as well as for the $P_w(E)$'s for $E_I=1$ and $E_I=0.02$ (which I have not displayed in graphical form and for which $d_{eff}$ takes on intermediate values). 
\begin{table}[h] 
\caption{The effective dimension ($d_{eff}^w$, from Equation~\eqref{eqn:deffdef}), evaluated for and averaged over 
the five sets of initial states used in this paper. The effective dimension is larger when the function $P(E)$ is broad and smooth. Comparing the curves in the top two panels of Figure~\ref{fig:P34forPaper} (as well as Figure~\ref{fig:P34forPaperZ}) suggests it is not surprising that $d_{eff}$ for $E_I=0.1$ is more than $60$ times greater than the $E_I=0.007$ case. (The quantity $\Delta$ gives the variance of $d_{eff}$ across the five solutions.) 
\label{tab:deffs}}
\begin{tabular}{c|c|c|c}
\toprule
\boldmath{$E_I$}	& \boldmath{$\left<d^w_{eff}\right>/N_w$}	& \boldmath{$\Delta$} & \boldmath{\% of $d^w_{eff}(E_I=0.1)$}\\
\colrule
1	& 0.087			& 30\%      &  27\% \\
0.1		& 0.24			& 8\% &  100\% \\
0.02	& 	0.06		& 33\% &  25\% \\
0.007	& 	0.004		& 70\% &  1.6\% \\
\colrule 
\end{tabular}
\end{table}

 The extremely different natures of the $E_I=0.1$ and $E_I=0.007$ curves are nicely captured by the large difference between their $d_{eff}$ values.  In addition, $\Delta$, the variance of $d_{eff}$ across the five different states gives one measure of ``scatter'' among the different $P_w(E)$ curves for fixed $E_I$.  This scatter is smallest for the $E_I=0.1$ case, which is consistent with the observations made about energy distributions in Section~\ref{sec:edist} (although those were focused on energy distributions in the subsystems).

\section{Tuning of States and Parameters}
\label{sec:tuning}
In the analysis presented here, the ``thermalized''-like behavior seems to emerge as a special case for a particular value (presumably actually a small region of values) for $E_I$. In much larger systems exhibiting localization discussed in the literature, it is typically the non-thermalized behavior that seems special, usually associated with specific parameter choices that lead to integrability.  I simply note here that it is not surprising that such matters of tuning depend on measures implicit in the model being considered. I regard the ACL model as too simplistic to draw broad conclusions about tuning of parameters, except as an illustration of how measures can turn out differently.  If the ``emergent laws'' perspective mentioned in the introduction is ever realized, that will come with its own perspective (and probably challenges) regarding measures. 

I also note that tuning of the initial state is involved in the notion of equilibration. A special choice of initial state with a low entropy is required in order to see a system dynamically approach equilibrium.  The fact that the actual Universe did indeed have such a special low entropy initial state is a source of great interest and curiosity to me, and although it is not often stated that way, it is related to the notorious ``tuning problems'' in cosmology (see~\cite{Penrose1979} for pioneering work and Section 5 of~\cite{ACLeqm} for a recent summary).  That is certainly not (directly) the topic of this paper, although I can not help but note with interest the very different perspective I sometimes see in the statistical mechanics literature (for example,~\cite{Popescu2006NatPh...2..754P} which implies that the Universe should be taken to be in a typical state).

\section{Discussion and Conclusions}
\label{sec:dc}
This research originated with my curiosity about various behaviors of the ACL model that I encountered in earlier work~\cite{ACLintro,ACLeqm}. On one hand, the equilibration process seemed so robust I wondered if there was a straightforward ergodicity picture to back it up. On the other hand, examination of the energy distributions that appeared in equilibrium made it clear that no conventional notion of temperature applied.  Furthermore, standard arguments would interpret the part of the environment density of states $N(E)$ that decreases with $E$ (see Figure~\ref{fig:NieOFtFirst}) as a ``negative temperature''.  Would that introduce strange artifacts in \mbox{our results?}

In this work, I have examined the equilibration processes in the ACL model systematically.  I have seen how the dephasing process is the solid foundation on which the equilibration takes place. Dephasing is able to drive equilibration under conditions where the notions of temperature and ergodicity do not apply. I have argued that this very basic form of equilibration is sufficient to support the use of the ACL model in studies of the equilibrium phenomena explored in~\cite{ACLeqm}.

Even though the notion of temperature does not apply, I have considered some primitive aspects of ``thermalization.''  Specifically, I have considered the expectation that different initial states with the same global energy thermalize to the same subsystem energy distributions. The ACL model is only able to realize this expectation in equilibrium for certain values of the coupling strength.  When this aspect of thermalization is realized, the energy distributions in the global space are smooth.  In the other cases, the global energy distribution can be quite jagged. I have related these different behaviors to the intrinsic properties of the global energy eigenstates, and argued that the smooth behavior could be viewed as something of a precursor to ergodicity, which might take a more concrete form in the limit of larger system sizes.  In addition, I have noted some rough parallels with discussions of the presence or absence of thermalization in large condensed matter~systems.   

Regarding negative temperature, even in the absence of a solid notion of temperature, the evolution of the energy distribution depicted in Figure~\ref{fig:NieOFtFirst} toward regions of lower energy but higher density of states might be seen as a more primitive version of the phenomena that can be associated with a negative temperature in other systems.

I have found it interesting to learn the degree to which the very simple ACL model is able to reflect certain familiar elements of equilibration, while still missing out on others due to its small size and other ``unphysical'' aspects.  Understanding systems such as this one that are on the edge of familiar behaviors could prove useful in exploring selection effects in frameworks where the laws of physics themselves are emergent, one of the motivations for this research I discussed in the Introduction.  Such work might ultimately help us understand the origin of the specific behaviors of the world around us that we \mbox{call ``physical''.}


\section{Reflections}
\label{sec:refl}
It is a great pleasure to contribute to this volume honoring Wojciech Zurek's 70th birthday. I first met Wojciech at an Aspen Center for Physics workshop the summer after I completed my PhD in 1983.  I have had the good fortune of having numerous connections with Wojciech since then, including as his postdoc later in the 1980s.  Wojciech has been an inspiration to me in many ways. For one, his unbounded and energetic curiosity has led to some of the most joyful and adventurous conversations of my entire career.  It is definitely in the spirit of this adventurous style that I have pursued the topics of this paper.  
I am also grateful to Wojciech for helping me develop a taste for natural hot springs.  It is fitting that certain advances on this project were made while partaking of some of my local favorites (experiencing temperature, but fortunately not equilibrium).

\section{Acknowledgements}
I am grateful to Rose Baunach, Zoe Holmes, Veronika Hubeny, Richard Scalettar and Rajiv Singh for helpful conversations.
This work was supported in part by the U.S. Department of Energy, Office of Science, Office of High Energy Physics QuantISED program under Contract No. KA2401032.

\appendix 
\section{Initial Conditions}
\label{app:IC}  
As discussed in~\cite{ACLintro}, coherent states can be constructed for the truncated SHO using the standard formula 
\begin{equation}
  \left|  \alpha 
 \right>  = \exp \left( {\alpha {{\bf{\hat a}}^\dag } - {\alpha ^*}{\bf{\hat a}}} \right)\left| 0 \right> .   
\end{equation}
The five initial states used throughout this paper are constructed by selecting the $i$th eigenstate of $H_e$, with $i$ drawn from $\left\{ 300,400,450, 500, 550 \right\} $ (ordered with increasing eigenvalue).  Recall that $N_e=600$. 

I then adjust $\alpha$ so that the initial state $\left|\alpha\right>_s\left|i\right>_e$ has $\left<H_w\right>=25$.  Note that $\left<H_w\right>$ includes the interaction term, so the value of $\alpha$ technically depends on $E_I$.  Except for the strongly coupled $E_I=1$ case, this dependence is very weak.  The exact values of the different initial subsystem energies in each case can be read from Figures~\ref{fig:AllFlows020}--\ref{fig:AllFlows1}. 

\section{Dephasing, Decoherence and Dissipation}
\label{app:DDD}
 {This work focuses} on dissipative processes which cause energy to flow and then stabilize at equilibrium values.  My focus on energy flow is natural for studying the topic of thermalization. In~\cite{ACLintro}, we explored how decoherence and dissipation happen on different time scales in the ACL model, a difference which is much more pronounced for larger systems.   In addition,  I note here that the term dephasing can mean different things, for example, relating the phases of two different beams as discussed in~\cite{RAUCH1999277}, where the notion of depolarization is also explored and contrasted. I use dephasing to refer to the randomization of the $\alpha^i_w$ phases over time. The dephasing I observe is not absolute, since recurrences are to be expected eventually.  However, the recurrence time lies far beyond the time ranges explored here (and in fact far outside the dynamic range of my computations, which are documented in~\cite{ACLintro}). This arrangement of the various relative time scales is well suited for the topics of this paper. 
 
\section{Eigenstate Thermalization Hypothesis}
\label{app:ETH}
The Eigenstate Thermalization Hypothesis (ETH)~\cite{Deutsch_PhysRevA.43.2046,Srednicki_1994,DAlessio_doi:10.1080/00018732.2016.1198134} proposes that important statistical properties of thermalized systems can be expressed by single eigenstates of the global Hamiltonian.  Even though the ETH is intended to apply to much larger systems, out of curiosity I have explored how the ACL behaviors might relate to the ETH. Focusing on the $P_s(E)$'s and $P_e(E)$'s, one can see from Figures~\ref{fig:Eigs_007_close_1} and~\ref{fig:Eigs_100_close_1} that the subsystem energy distributions look more similar for individual eigenstates of $H_w$ for $E_I=0.1$ than for $E_I=0.007$.  This similarity suggests that for the ``thermalized'' $E_I=0.1$ case individual eigenstates are starting to line up behind a common statement about the subsystem energy distributions, which would be a signature of the ETH. But the fact that significant differences remain among the $E_I=0.1$ curves suggests this signature is not expressed strongly. 

\begin{figure}[h]
\includegraphics[width=3.7 in]{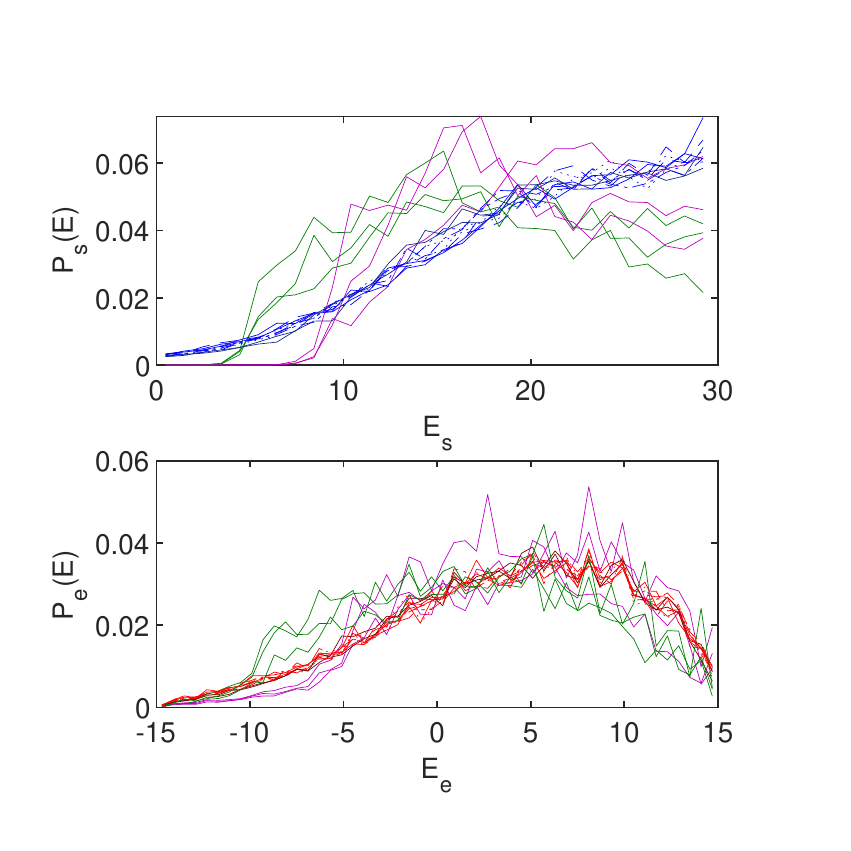}
\caption{Exploring ETH:  The energy distributions for the $E_I = 0.1$ case from Figures~\ref{fig:NisLateSetMultiEi} (system) and~\ref{fig:NieLateSetMultiEi} (environment) are shown (blue and red curves, respectively).  In addition, the energy distributions for individual eigenstates of $H_w$ are plotted.  As depicted (and color coded) in Figure~\ref{fig:ETHeigs}, the particular $H_w$ eigenstates are chosen to be representative of the global energy distributions that correspond to the original curves from Figures~\ref{fig:NisLateSetMultiEi} and~\ref{fig:NieLateSetMultiEi}.   No single eigenstate fully characterizes the equilibrium distributions (although they are not completely off the mark), and there is significant scatter among the distributions drawn from the individual eigenstates.  These results suggest only a very loose realization of the ETH. 
\label{fig:ETHdists_es}}
\end{figure}

\begin{figure}[h]
\includegraphics[width=3.7 in]{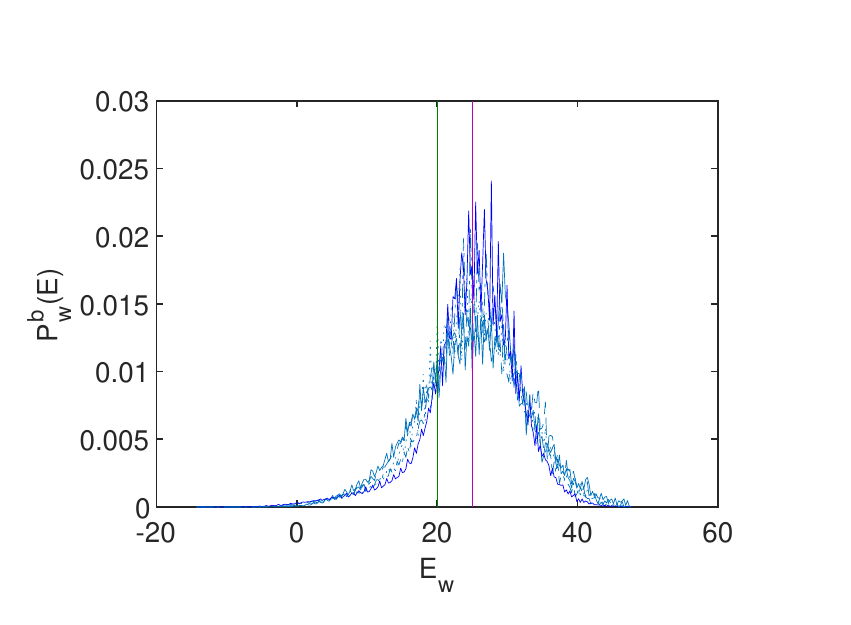}
\caption{Selected eigenstates:  The eigenvalues of the $H_w$ eigenstates chosen for Figure~\ref{fig:ETHdists_es} are shown here (vertical lines), against the backdrop of the global energy distributions from the $E_I=0.1$ panel in Figure~\ref{fig:P34forPaper}, which correspond to the original energy distributions depicted for the subsystems in Figure~\ref{fig:ETHdists_es}.  There are three green and three purple lines, which are unresolved in the figure because they show adjacent eigenvalues (which only differ at the $0.01\%$ level).  The curves in Figure~\ref{fig:ETHdists_es} are matched by color with the eigenvalues shown here. 
\label{fig:ETHeigs}}
\end{figure}

Figure~\ref{fig:ETHdists_es} presents a more systematic analysis of the $H_w$ eigenstates relevant to the sets of initial conditions used here. 
(The selection of the eigenstates is illustrated in Figure~\ref{fig:ETHeigs}.)
The $P_s(E)$ and $P_e(E)$ curves for those eigenstates approach the energy distributions for our set of initial states reasonably well, but they still retain considerably greater scatter than seen for the non-$H_w$ eigenstate states. Thus, the averaging over the entire distribution in $P_w(E)$ space appears to have a significant role in realizing the sharpness of the convergence of the $P_s(E)$ and $P_e(E)$ curves.  Although one could try to argue that there are elements of the ETH being expressed here, any such expression is only an approximate one.  

\bibliography{ThisOne}

\begin{thebibliography}{22}%
\makeatletter
\providecommand \@ifxundefined [1]{%
 \@ifx{#1\undefined}
}%
\providecommand \@ifnum [1]{%
 \ifnum #1\expandafter \@firstoftwo
 \else \expandafter \@secondoftwo
 \fi
}%
\providecommand \@ifx [1]{%
 \ifx #1\expandafter \@firstoftwo
 \else \expandafter \@secondoftwo
 \fi
}%
\providecommand \natexlab [1]{#1}%
\providecommand \enquote  [1]{``#1''}%
\providecommand \bibnamefont  [1]{#1}%
\providecommand \bibfnamefont [1]{#1}%
\providecommand \citenamefont [1]{#1}%
\providecommand \href@noop [0]{\@secondoftwo}%
\providecommand \href [0]{\begingroup \@sanitize@url \@href}%
\providecommand \@href[1]{\@@startlink{#1}\@@href}%
\providecommand \@@href[1]{\endgroup#1\@@endlink}%
\providecommand \@sanitize@url [0]{\catcode `\\12\catcode `\$12\catcode
  `\&12\catcode `\#12\catcode `\^12\catcode `\_12\catcode `\%12\relax}%
\providecommand \@@startlink[1]{}%
\providecommand \@@endlink[0]{}%
\providecommand \url  [0]{\begingroup\@sanitize@url \@url }%
\providecommand \@url [1]{\endgroup\@href {#1}{\urlprefix }}%
\providecommand \urlprefix  [0]{URL }%
\providecommand \Eprint [0]{\href }%
\providecommand \doibase [0]{https://doi.org/}%
\providecommand \selectlanguage [0]{\@gobble}%
\providecommand \bibinfo  [0]{\@secondoftwo}%
\providecommand \bibfield  [0]{\@secondoftwo}%
\providecommand \translation [1]{[#1]}%
\providecommand \BibitemOpen [0]{}%
\providecommand \bibitemStop [0]{}%
\providecommand \bibitemNoStop [0]{.\EOS\space}%
\providecommand \EOS [0]{\spacefactor3000\relax}%
\providecommand \BibitemShut  [1]{\csname bibitem#1\endcsname}%
\let\auto@bib@innerbib\@empty
\bibitem [{\citenamefont {Albrecht}\ \emph
  {et~al.}(2021{\natexlab{a}})\citenamefont {Albrecht}, \citenamefont
  {Baunach},\ and\ \citenamefont {Arrasmith}}]{ACLintro}%
  \BibitemOpen
  \bibfield  {author} {\bibinfo {author} {\bibfnamefont {A.}~\bibnamefont
  {Albrecht}}, \bibinfo {author} {\bibfnamefont {R.}~\bibnamefont {Baunach}},\
  and\ \bibinfo {author} {\bibfnamefont {A.}~\bibnamefont {Arrasmith}},\
  }\href@noop {} {\bibinfo {title} {{Adapted Caldeira-Leggett Model}}}
  (\bibinfo {year} {2021}{\natexlab{a}}),\ \Eprint
  {https://arxiv.org/abs/2105.14040} {arXiv:2105.14040 [quant-ph]} \BibitemShut
  {NoStop}%
\bibitem [{\citenamefont {Albrecht}\ \emph
  {et~al.}(2021{\natexlab{b}})\citenamefont {Albrecht}, \citenamefont
  {Baunach},\ and\ \citenamefont {Arrasmith}}]{ACLeqm}%
  \BibitemOpen
  \bibfield  {author} {\bibinfo {author} {\bibfnamefont {A.}~\bibnamefont
  {Albrecht}}, \bibinfo {author} {\bibfnamefont {R.}~\bibnamefont {Baunach}},\
  and\ \bibinfo {author} {\bibfnamefont {A.}~\bibnamefont {Arrasmith}},\
  }\href@noop {} {\bibinfo {title} {Einselection, equilibrium and cosmology}}
  (\bibinfo {year} {2021}{\natexlab{b}}),\ \Eprint
  {https://arxiv.org/abs/2105.14017} {arXiv:2105.14017 [hep-th]} \BibitemShut
  {NoStop}%
\bibitem [{\citenamefont {Zurek}\ \emph {et~al.}(1993)\citenamefont {Zurek},
  \citenamefont {Habib},\ and\ \citenamefont {Paz}}]{Zurek:1992mv}%
  \BibitemOpen
  \bibfield  {author} {\bibinfo {author} {\bibfnamefont {W.~H.}\ \bibnamefont
  {Zurek}}, \bibinfo {author} {\bibfnamefont {S.}~\bibnamefont {Habib}},\ and\
  \bibinfo {author} {\bibfnamefont {J.~P.}\ \bibnamefont {Paz}},\ }\bibfield
  {title} {\bibinfo {title} {{Coherent states via decoherence}},\ }\href
  {https://doi.org/10.1103/PhysRevLett.70.1187} {\bibfield  {journal} {\bibinfo
   {journal} {Phys. Rev. Lett.}\ }\textbf {\bibinfo {volume} {70}},\ \bibinfo
  {pages} {1187} (\bibinfo {year} {1993})}\BibitemShut {NoStop}%
\bibitem [{\citenamefont {Baunach}\ \emph {et~al.}(2021)\citenamefont
  {Baunach}, \citenamefont {Albrecht},\ and\ \citenamefont
  {Arrasmith}}]{CopyCat}%
  \BibitemOpen
  \bibfield  {author} {\bibinfo {author} {\bibfnamefont {R.}~\bibnamefont
  {Baunach}}, \bibinfo {author} {\bibfnamefont {A.}~\bibnamefont {Albrecht}},\
  and\ \bibinfo {author} {\bibfnamefont {A.}~\bibnamefont {Arrasmith}},\
  }\href@noop {} {\bibinfo {title} {Copycat process in the early stages of
  einselection}} (\bibinfo {year} {2021}),\ \Eprint
  {https://arxiv.org/abs/2105.14032} {arXiv:2105.14032 [quant-ph]} \BibitemShut
  {NoStop}%
\bibitem [{\citenamefont {Albrecht}\ and\ \citenamefont
  {Iglesias}(2008)}]{Albrecht:2007mm}%
  \BibitemOpen
  \bibfield  {author} {\bibinfo {author} {\bibfnamefont {A.}~\bibnamefont
  {Albrecht}}\ and\ \bibinfo {author} {\bibfnamefont {A.}~\bibnamefont
  {Iglesias}},\ }\bibfield  {title} {\bibinfo {title} {{The Clock ambiguity and
  the emergence of physical laws}},\ }\href
  {https://doi.org/10.1103/PhysRevD.77.063506} {\bibfield  {journal} {\bibinfo
  {journal} {Phys. Rev. D}\ }\textbf {\bibinfo {volume} {77}},\ \bibinfo
  {pages} {063506} (\bibinfo {year} {2008})},\ \Eprint
  {https://arxiv.org/abs/0708.2743} {arXiv:0708.2743 [hep-th]} \BibitemShut
  {NoStop}%
\bibitem [{\citenamefont {Albrecht}\ and\ \citenamefont
  {Iglesias}(2015)}]{AlbrechtPhysRevD.91.043529}%
  \BibitemOpen
  \bibfield  {author} {\bibinfo {author} {\bibfnamefont {A.}~\bibnamefont
  {Albrecht}}\ and\ \bibinfo {author} {\bibfnamefont {A.}~\bibnamefont
  {Iglesias}},\ }\bibfield  {title} {\bibinfo {title} {Lorentz symmetric
  dispersion relation from a random hamiltonian},\ }\href
  {https://doi.org/10.1103/PhysRevD.91.043529} {\bibfield  {journal} {\bibinfo
  {journal} {Phys. Rev. D}\ }\textbf {\bibinfo {volume} {91}},\ \bibinfo
  {pages} {043529} (\bibinfo {year} {2015})}\BibitemShut {NoStop}%
\bibitem [{\citenamefont {Binder}\ \emph {et~al.}(2019)\citenamefont {Binder},
  \citenamefont {Correa}, \citenamefont {Gogolin}, \citenamefont {Anders},\
  and\ \citenamefont {Adesso}}]{binder2019thermodynamics}%
  \BibitemOpen
  \bibfield  {author} {\bibinfo {author} {\bibfnamefont {F.}~\bibnamefont
  {Binder}}, \bibinfo {author} {\bibfnamefont {L.}~\bibnamefont {Correa}},
  \bibinfo {author} {\bibfnamefont {C.}~\bibnamefont {Gogolin}}, \bibinfo
  {author} {\bibfnamefont {J.}~\bibnamefont {Anders}},\ and\ \bibinfo {author}
  {\bibfnamefont {G.}~\bibnamefont {Adesso}},\ }\href
  {https://books.google.com/books?id=5uWPDwAAQBAJ} {\emph {\bibinfo {title}
  {Thermodynamics in the Quantum Regime: Fundamental Aspects and New
  Directions}}},\ Fundamental Theories of Physics\ (\bibinfo  {publisher}
  {Springer International Publishing},\ \bibinfo {year} {2019})\BibitemShut
  {NoStop}%
\bibitem [{\citenamefont {Lloyd}(2006)}]{Lloyd2006ExcuseOI}%
  \BibitemOpen
  \bibfield  {author} {\bibinfo {author} {\bibfnamefont {S.}~\bibnamefont
  {Lloyd}},\ }\bibfield  {title} {\bibinfo {title} {Excuse our ignorance},\
  }\href@noop {} {\bibfield  {journal} {\bibinfo  {journal} {Nature Physics}\
  }\textbf {\bibinfo {volume} {2}},\ \bibinfo {pages} {727} (\bibinfo {year}
  {2006})}\BibitemShut {NoStop}%
\bibitem [{\citenamefont {Cucchietti}\ \emph {et~al.}(2005)\citenamefont
  {Cucchietti}, \citenamefont {Paz},\ and\ \citenamefont {Zurek}}]{CPZ1}%
  \BibitemOpen
  \bibfield  {author} {\bibinfo {author} {\bibfnamefont {F.~M.}\ \bibnamefont
  {Cucchietti}}, \bibinfo {author} {\bibfnamefont {J.~P.}\ \bibnamefont
  {Paz}},\ and\ \bibinfo {author} {\bibfnamefont {W.~H.}\ \bibnamefont
  {Zurek}},\ }\bibfield  {title} {\bibinfo {title} {Decoherence from spin
  environments},\ }\href {https://doi.org/10.1103/PhysRevA.72.052113}
  {\bibfield  {journal} {\bibinfo  {journal} {Phys. Rev. A}\ }\textbf {\bibinfo
  {volume} {72}},\ \bibinfo {pages} {052113} (\bibinfo {year}
  {2005})}\BibitemShut {NoStop}%
\bibitem [{\citenamefont {Zurek}\ \emph {et~al.}(2006)\citenamefont {Zurek},
  \citenamefont {Cucchietti},\ and\ \citenamefont {Paz}}]{CPZ2}%
  \BibitemOpen
  \bibfield  {author} {\bibinfo {author} {\bibfnamefont {W.~H.}\ \bibnamefont
  {Zurek}}, \bibinfo {author} {\bibfnamefont {F.~M.}\ \bibnamefont
  {Cucchietti}},\ and\ \bibinfo {author} {\bibfnamefont {J.~P.}\ \bibnamefont
  {Paz}},\ }\href@noop {} {\bibinfo {title} {Gaussian decoherence and gaussian
  echo from spin environments}} (\bibinfo {year} {2006}),\ \Eprint
  {https://arxiv.org/abs/quant-ph/0611200} {arXiv:quant-ph/0611200 [quant-ph]}
  \BibitemShut {NoStop}%
\bibitem [{\citenamefont {Nandkishore}\ and\ \citenamefont
  {Huse}(2015)}]{Nandkishoredoi:10.1146/annurev-conmatphys-031214-014726}%
  \BibitemOpen
  \bibfield  {author} {\bibinfo {author} {\bibfnamefont {R.}~\bibnamefont
  {Nandkishore}}\ and\ \bibinfo {author} {\bibfnamefont {D.~A.}\ \bibnamefont
  {Huse}},\ }\bibfield  {title} {\bibinfo {title} {Many-body localization and
  thermalization in quantum statistical mechanics},\ }\href
  {https://doi.org/10.1146/annurev-conmatphys-031214-014726} {\bibfield
  {journal} {\bibinfo  {journal} {Annual Review of Condensed Matter Physics}\
  }\textbf {\bibinfo {volume} {6}},\ \bibinfo {pages} {15} (\bibinfo {year}
  {2015})},\ \Eprint
  {https://arxiv.org/abs/https://doi.org/10.1146/annurev-conmatphys-031214-014726}
  {https://doi.org/10.1146/annurev-conmatphys-031214-014726} \BibitemShut
  {NoStop}%
\bibitem [{\citenamefont {{Popescu}}\ \emph {et~al.}(2006)\citenamefont
  {{Popescu}}, \citenamefont {{Short}},\ and\ \citenamefont
  {{Winter}}}]{Popescu2006NatPh...2..754P}%
  \BibitemOpen
  \bibfield  {author} {\bibinfo {author} {\bibfnamefont {S.}~\bibnamefont
  {{Popescu}}}, \bibinfo {author} {\bibfnamefont {A.~J.}\ \bibnamefont
  {{Short}}},\ and\ \bibinfo {author} {\bibfnamefont {A.}~\bibnamefont
  {{Winter}}},\ }\bibfield  {title} {\bibinfo {title} {{Entanglement and the
  foundations of statistical mechanics}},\ }\href
  {https://doi.org/10.1038/nphys444} {\bibfield  {journal} {\bibinfo  {journal}
  {Nature Physics}\ }\textbf {\bibinfo {volume} {2}},\ \bibinfo {pages} {754}
  (\bibinfo {year} {2006})},\ \Eprint {https://arxiv.org/abs/quant-ph/0511225}
  {arXiv:quant-ph/0511225 [quant-ph]} \BibitemShut {NoStop}%
\bibitem [{\citenamefont {Rigol}\ \emph {et~al.}(2007)\citenamefont {Rigol},
  \citenamefont {Dunjko}, \citenamefont {Yurovsky},\ and\ \citenamefont
  {Olshanii}}]{RigolPhysRevLett.98.050405}%
  \BibitemOpen
  \bibfield  {author} {\bibinfo {author} {\bibfnamefont {M.}~\bibnamefont
  {Rigol}}, \bibinfo {author} {\bibfnamefont {V.}~\bibnamefont {Dunjko}},
  \bibinfo {author} {\bibfnamefont {V.}~\bibnamefont {Yurovsky}},\ and\
  \bibinfo {author} {\bibfnamefont {M.}~\bibnamefont {Olshanii}},\ }\bibfield
  {title} {\bibinfo {title} {Relaxation in a completely integrable many-body
  quantum system: An ab initio study of the dynamics of the highly excited
  states of 1d lattice hard-core bosons},\ }\href
  {https://doi.org/10.1103/PhysRevLett.98.050405} {\bibfield  {journal}
  {\bibinfo  {journal} {Phys. Rev. Lett.}\ }\textbf {\bibinfo {volume} {98}},\
  \bibinfo {pages} {050405} (\bibinfo {year} {2007})}\BibitemShut {NoStop}%
\bibitem [{\citenamefont {Lenard}(1978)}]{lenard1978thermodynamical}%
  \BibitemOpen
  \bibfield  {author} {\bibinfo {author} {\bibfnamefont {A.}~\bibnamefont
  {Lenard}},\ }\bibfield  {title} {\bibinfo {title} {Thermodynamical proof of
  the gibbs formula for elementary quantum systems},\ }\href@noop {} {\bibfield
   {journal} {\bibinfo  {journal} {Journal of Statistical Physics}\ }\textbf
  {\bibinfo {volume} {19}},\ \bibinfo {pages} {575} (\bibinfo {year}
  {1978})}\BibitemShut {NoStop}%
\bibitem [{\citenamefont {Ikeda}\ \emph {et~al.}(2011)\citenamefont {Ikeda},
  \citenamefont {Watanabe},\ and\ \citenamefont
  {Ueda}}]{IkedaPhysRevE.84.021130}%
  \BibitemOpen
  \bibfield  {author} {\bibinfo {author} {\bibfnamefont {T.~N.}\ \bibnamefont
  {Ikeda}}, \bibinfo {author} {\bibfnamefont {Y.}~\bibnamefont {Watanabe}},\
  and\ \bibinfo {author} {\bibfnamefont {M.}~\bibnamefont {Ueda}},\ }\bibfield
  {title} {\bibinfo {title} {Eigenstate randomization hypothesis: Why does the
  long-time average equal the microcanonical average?},\ }\href
  {https://doi.org/10.1103/PhysRevE.84.021130} {\bibfield  {journal} {\bibinfo
  {journal} {Phys. Rev. E}\ }\textbf {\bibinfo {volume} {84}},\ \bibinfo
  {pages} {021130} (\bibinfo {year} {2011})}\BibitemShut {NoStop}%
\bibitem [{\citenamefont {Paz}\ and\ \citenamefont {Zurek}(1999)}]{Paz:1998ib}%
  \BibitemOpen
  \bibfield  {author} {\bibinfo {author} {\bibfnamefont {J.~P.}\ \bibnamefont
  {Paz}}\ and\ \bibinfo {author} {\bibfnamefont {W.~H.}\ \bibnamefont
  {Zurek}},\ }\bibfield  {title} {\bibinfo {title} {{Quantum limit of
  decoherence: Environment induced superselection of energy eigenstates}},\
  }\href {https://doi.org/10.1103/PhysRevLett.82.5181} {\bibfield  {journal}
  {\bibinfo  {journal} {Phys. Rev. Lett.}\ }\textbf {\bibinfo {volume} {82}},\
  \bibinfo {pages} {5181} (\bibinfo {year} {1999})},\ \Eprint
  {https://arxiv.org/abs/quant-ph/9811026} {arXiv:quant-ph/9811026}
  \BibitemShut {NoStop}%
\bibitem [{\citenamefont {Serbyn}\ \emph {et~al.}(2013)\citenamefont {Serbyn},
  \citenamefont {Papi\ifmmode~\acute{c}\else \'{c}\fi{}},\ and\ \citenamefont
  {Abanin}}]{SerbynPhysRevLett.111.127201}%
  \BibitemOpen
  \bibfield  {author} {\bibinfo {author} {\bibfnamefont {M.}~\bibnamefont
  {Serbyn}}, \bibinfo {author} {\bibfnamefont {Z.}~\bibnamefont
  {Papi\ifmmode~\acute{c}\else \'{c}\fi{}}},\ and\ \bibinfo {author}
  {\bibfnamefont {D.~A.}\ \bibnamefont {Abanin}},\ }\bibfield  {title}
  {\bibinfo {title} {Local conservation laws and the structure of the many-body
  localized states},\ }\href {https://doi.org/10.1103/PhysRevLett.111.127201}
  {\bibfield  {journal} {\bibinfo  {journal} {Phys. Rev. Lett.}\ }\textbf
  {\bibinfo {volume} {111}},\ \bibinfo {pages} {127201} (\bibinfo {year}
  {2013})}\BibitemShut {NoStop}%
\bibitem [{\citenamefont {Penrose}(1979)}]{Penrose1979}%
  \BibitemOpen
  \bibfield  {author} {\bibinfo {author} {\bibfnamefont {R.}~\bibnamefont
  {Penrose}},\ }\bibfield  {title} {\bibinfo {title} {{Singularities and
  time-asymmetry}},\ }in\ \href@noop {} {\emph {\bibinfo {booktitle} {General
  {R}elativity, an {E}instein {C}entenary {S}urvey}}}\ (\bibinfo  {publisher}
  {Cambridge},\ \bibinfo {year} {1979})\BibitemShut {NoStop}%
\bibitem [{\citenamefont {Rauch}\ \emph {et~al.}(1999)\citenamefont {Rauch},
  \citenamefont {Suda},\ and\ \citenamefont {Pascazio}}]{RAUCH1999277}%
  \BibitemOpen
  \bibfield  {author} {\bibinfo {author} {\bibfnamefont {H.}~\bibnamefont
  {Rauch}}, \bibinfo {author} {\bibfnamefont {M.}~\bibnamefont {Suda}},\ and\
  \bibinfo {author} {\bibfnamefont {S.}~\bibnamefont {Pascazio}},\ }\bibfield
  {title} {\bibinfo {title} {Decoherence, dephasing and depolarization},\
  }\href {https://doi.org/https://doi.org/10.1016/S0921-4526(99)00082-4}
  {\bibfield  {journal} {\bibinfo  {journal} {Physica B: Condensed Matter}\
  }\textbf {\bibinfo {volume} {267-268}},\ \bibinfo {pages} {277} (\bibinfo
  {year} {1999})}\BibitemShut {NoStop}%
\bibitem [{\citenamefont {Deutsch}(1991)}]{Deutsch_PhysRevA.43.2046}%
  \BibitemOpen
  \bibfield  {author} {\bibinfo {author} {\bibfnamefont {J.~M.}\ \bibnamefont
  {Deutsch}},\ }\bibfield  {title} {\bibinfo {title} {Quantum statistical
  mechanics in a closed system},\ }\href
  {https://doi.org/10.1103/PhysRevA.43.2046} {\bibfield  {journal} {\bibinfo
  {journal} {Phys. Rev. A}\ }\textbf {\bibinfo {volume} {43}},\ \bibinfo
  {pages} {2046} (\bibinfo {year} {1991})}\BibitemShut {NoStop}%
\bibitem [{\citenamefont {Srednicki}(1994)}]{Srednicki_1994}%
  \BibitemOpen
  \bibfield  {author} {\bibinfo {author} {\bibfnamefont {M.}~\bibnamefont
  {Srednicki}},\ }\bibfield  {title} {\bibinfo {title} {Chaos and quantum
  thermalization},\ }\href {https://doi.org/10.1103/physreve.50.888} {\bibfield
   {journal} {\bibinfo  {journal} {Physical Review E}\ }\textbf {\bibinfo
  {volume} {50}},\ \bibinfo {pages} {888–901} (\bibinfo {year}
  {1994})}\BibitemShut {NoStop}%
\bibitem [{\citenamefont {D'Alessio}\ \emph {et~al.}(2016)\citenamefont
  {D'Alessio}, \citenamefont {Kafri}, \citenamefont {Polkovnikov},\ and\
  \citenamefont {Rigol}}]{DAlessio_doi:10.1080/00018732.2016.1198134}%
  \BibitemOpen
  \bibfield  {author} {\bibinfo {author} {\bibfnamefont {L.}~\bibnamefont
  {D'Alessio}}, \bibinfo {author} {\bibfnamefont {Y.}~\bibnamefont {Kafri}},
  \bibinfo {author} {\bibfnamefont {A.}~\bibnamefont {Polkovnikov}},\ and\
  \bibinfo {author} {\bibfnamefont {M.}~\bibnamefont {Rigol}},\ }\bibfield
  {title} {\bibinfo {title} {From quantum chaos and eigenstate thermalization
  to statistical mechanics and thermodynamics},\ }\href
  {https://doi.org/10.1080/00018732.2016.1198134} {\bibfield  {journal}
  {\bibinfo  {journal} {Advances in Physics}\ }\textbf {\bibinfo {volume}
  {65}},\ \bibinfo {pages} {239} (\bibinfo {year} {2016})},\ \Eprint
  {https://arxiv.org/abs/https://doi.org/10.1080/00018732.2016.1198134}
  {https://doi.org/10.1080/00018732.2016.1198134} \BibitemShut {NoStop}%
\end{thebibliography}%

\end{document}